\documentclass[twocolumn,appendixfloats]{emulateapj}
\usepackage{float}
\usepackage{epsfig}
\usepackage{graphicx}
\usepackage{graphics}
\usepackage[latin1]{inputenc}
\usepackage{latexsym}
\newcommand{\nd}{\multicolumn{1}{c}{$\dots$}}

\newcommand{\ngal}{NGC$\,$4258}
\newcommand{\ngl}{N4258}
\newcommand{\mm}{\!-\!}
\newcommand{\ig}{\!=\!}
\newcommand{\ho}{H_0}
\newcommand{\nvar}{4143}
\newcommand{\gfrac}{54}
\newcommand{\nceph}{94}
\newcommand{\nvars}{215}
\newcommand{\nlsst}{77}
\slugcomment{Accepted for publication in the Astronomical Journal}
\shorttitle{Cepheids in \ngal}
\shortauthors{Hoffmann \& Macri}

\begin{document}

\title{Cepheid Variables in the Maser-Host Galaxy NGC$\,$4258}

\author{Samantha L.~Hoffmann \& Lucas M.~Macri$^{*}$}
\affil{George P. and Cynthia Woods Mitchell Institute for Fundamental Physics and Astronomy,\\Department of Physics and Astronomy, Texas A\&M University, College Station, TX 77843, USA}
\altaffiltext{*}{Corresponding author; {\tt lmacri@tamu.edu}}

\begin{abstract}
We present results of a ground-based survey for Cepheid variables in NGC$\,$4258. This galaxy plays a key role in the Extragalactic Distance Scale due to its very precise and accurate distance determination via VLBI observations of water masers. We imaged two fields within this galaxy using the Gemini North telescope and GMOS, obtaining 16 epochs of data in the SDSS {\it gri} bands over 4 years. We carried out PSF photometry and detected \nceph\ Cepheids with periods between 7 and 127 days, as well as an additional \nvars\ variables which may be Cepheids or Population II pulsators. We used the Cepheid sample to test the absolute calibration of theoretical {\it gri} Period-Luminosity relations and found good agreement with the maser distance to this galaxy. The expected data products from the Large Synoptic Survey Telescope (LSST) should enable Cepheid searches out to at least 10 Mpc.
\end{abstract}


\section{Introduction\label{sc:intro}}

The classical Extragalactic Distance Scale plays a key role in the current era of ``precision cosmology'' by providing an estimate of the Hubble constant ($\ho$) free from assumptions about the contents of our Universe. Hence, comparing the value of $\ho$ obtained via Cepheids and type Ia supernovae \citep[e.g.,][]{riess11} with the one inferred from BAO and CMB observations \citep{anderson14,planck13} can provide a strong additional constraint on the properties of dark energy and other cosmological parameters \citep{weinberg13,dvorkin14}.

\vspace{3pt}

\ngal\ is a critical anchor in the Cosmic Distance Ladder thanks to its very precise and accurate distance estimate based on VLBI observations of water masers orbiting its central massive black hole \citep{miyoshi95,herrnstein99,argon07,humphreys08}, with a current value of $D\ig 7.6\pm3$\%\,Mpc \citep[][equivalent to a distance modulus of $\mu_0\ig 29.404\pm0.065$~mag]{humphreys13}. It was previously surveyed for Cepheids by \citet{macri06}, who used the {\it Hubble Space Telescope} Advanced Camera for Surveys (ACS) to discover 281 variables with periods between 4 and 45 days. Recently, \citet{fausnaugh14} used the Large Binocular Telescope to survey the entire disk of \ngal\ for Cepheids and found 81 Cepheids with $13\!<P\!<90$~d. They used the technique developed by \citet{gerke11}, in which Cepheids are detected via difference-imaging of ground-based data and the photometric calibration is obtained from {\it Hubble} images.

\vspace{3pt}

Given the importance of \ngal\ for the Extragalactic Distance Scale, we wished to increase its sample of Cepheids with an emphasis on long-period objects. Among the 117 \ngal\ Cepheids used by \citet{riess11}, only 24\% have $P\!>\!30$d (11\% for $P\!>\!40$d), whereas the samples in the 8 SNe Ia hosts used in that work contain 72\% and 47\% of the objects in the same period ranges. A better match in the period range spanned by calibrator and target galaxies helps to decrease the impact of the systematic uncertainty associated with possible changes in the slope of the Cepheid Period-Luminosity (P-L) relation from galaxy to galaxy. An additional motivation for our study was to provide an empirical absolute calibration of the Cepheid Period-Luminosity relations in the SDSS {\it gri} filters, to supplement the semi-empirical approach of \citet{ngeow07} and the theoretical models of \citet{dicriscienzo13}. The use of this filter set for Cepheid work will become more prevalent in the era of LSST.

\vspace{3pt}

The rest of this paper is organized as follows: \S\ref{SC:OBS} presents the details of the observations; \S\ref{SC:DRPHOT} describes the data reduction, photometry and calibration; \S\ref{SC:SLOANPL} discusses the fiducial Cepheid P-L relations in the SDSS filters; \S\ref{SC:IDCEPH} details the procedures used to identify and classify Cepheid variables; \S\ref{SC:RES} discusses our results; and \S\ref{SC:LSST} explores the use of LSST for extragalactic Cepheid work.

\section{Observations\label{sc:obs}}

\subsection{Gemini North \label{sc:gemobs}}
We conducted the Cepheid search using the Gemini North 8.1-m telescope and the Gemini Multi-Object Spectrograph \citep[GMOS,][]{davies97}, under programs GN-2004A-Q-22 and GN-2007A-Q-14. GMOS has a $5\farcm5\times5\farcm5$ field of view that is covered by three charged-coupled devices (CCDs) with a scale of $0\farcs0727/$pixel. There are two small ($2\farcs8$) gaps between the CCDs and the corners of the outer chips are not illuminated.

\vspace{3pt}

\ngal\ was imaged on 22 nights over 4 years in order to ensure good phase coverage of the Cepheids. We targeted two fields within the galaxy located at different galactocentric distances, placed so they would fully contain the regions previously observed by \citet{macri06}. The GMOS field of view is $\sim 3\times$ that of ACS, so this overlap enables the recovery of long-period Cepheids previously discovered with HST while still significantly extending the area of the disk that is monitored for variables. \hspace{2pt} We \hspace{2pt} follow \hspace{2pt} the \hspace{2pt} naming \hspace{2pt} convention \hspace{2pt} adopted \par

\begin{deluxetable}{ll}
\tablecaption{Observation Log\label{tab:log}}
\tabletypesize{\scriptsize}
\tablewidth{0pc}
\tablehead{\colhead{Date} & \colhead{Images}}
\startdata
2004 Feb 18 & $g\times2$, $r\times2$, $i\times2$ (I,O)     \\
2004 Feb 20 & $g\times2$ (I,O); $r\times2$, $i\times2$ (I) \\
2004 Mar 29 & $g\times2$ (O) \\
2004 May 22 & $g\times2$ (I),$ \times$3 (O)\\
            & $r\times2$, $i\times2$ (I,O) \\
2004 May 24 & $g\times2$, $r\times2$, $i\times2$ (I,O) \\
2004 Jul 08 & $g\times2$, $r\times2$, $i\times2$ (I) \\
2004 Jul 14 & $r\times1$, $i\times2$ (O) \\
2005 Feb 10 & $g\times2$, $r\times2$, $i\times2$ (I,O) \\
2005 Mar 09 & $g\times2$, $r\times2$, $i\times2$ (I,O) \\
2005 Apr 09 & $g\times2$, $r\times2$, $i\times2$ (I,O) \\
2005 Apr 12 & $g\times2$, $r\times2$, $i\times2$ (I,O) \\
2005 May 04 & $g\times2$, $r\times2$, $i\times2$ (I,O) \\
2005 May 08 & $g\times2$, $r\times2$, $i\times2$ (I,O) \\
2007 Feb 22 & $g\times2$, $r\times2$, $i\times2$ (I,O) \\
2007 Apr 07 & $r\times2$, $i\times1$ (O) \\
2007 Apr 12 & $g\times2$, $r\times2$, $i\times2$ (I,O) \\
2007 Apr 20 & $g\times2$, $r\times2$, $i\times2$ (O) \\
2008 Jan 06 & $g\times2$, $r\times2$, $i\times2$ (I) \\
2008 Jan 07 & $g\times2$, $r\times2$, $i\times2$ (O) \\
2008 Jan 10 & $g\times2$, $r\times2$, $i\times2$ (I) \\
2008 Jan 16 & $g\times3$, $r\times2$, $i\times2$ (O) \\
2008 Feb 16 & $g\times2$, $r\times2$, $i\times2$ (I)
\enddata
\tablecomments{I: inner field; O: outer field.}
\end{deluxetable}

\ \par
\vspace{6pt}

\noindent{by \citet{macri06}: the field located at a larger galactocentric distance is called ``outer'' and the one closer to the galaxy nucleus is called ``inner''. The GMOS fields were centered at $\alpha\ig 12^{\rm h}19^{\rm m}20.16^{\rm s}, \delta\ig +47^{\circ}12\arcmin33\farcs3$ and $\alpha\ig 12^{\rm h}18^{\rm m}48.21^{\rm s}, \delta\ig +47^{\circ}20\arcmin25\farcs8$ (J2000.0) for the ``outer'' and ``inner'' fields respectively. Figure~\ref{fig:mos} shows the location of the fields within the galaxy.}
\vspace{3pt}

We typically obtained $2\times 600$~s exposures at each epoch using the Sloan Digital Sky Survey (SDSS) {\it gri} filters \citep{fukugita96}. The observations were obtained in queue mode by Gemini staff when the sky conditions were clear (although not necessarily photometric) and the seeing was below $0\farcs7$; 16 useful epochs were obtained for each field. The observation log is presented in Table~\ref{tab:log}. 

\subsection{WIYN \label{sc:wiynobs}} 
\begin{figure}[htbp]
\begin{center}
\includegraphics[width=0.49\textwidth]{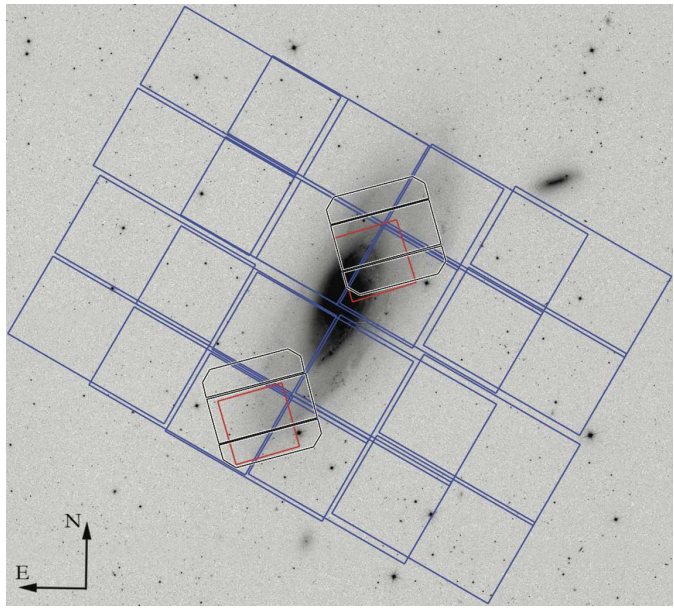}
\end{center}
\vspace{-6pt}
\caption{SDSS $r$-band mosaic ($36\arcmin\times32\arcmin$) of \ngal\ showing the footprints of the GMOS (octagonal, black and white), WIYN (square, blue) and \citet{macri06} HST/ACS (square, red) fields. The ``inner'' field is located north of the galaxy center, while the ``outer'' field is located south and east. \label{fig:mos}}
\end{figure}

In order to perform a photometric calibration of the Gemini data (see \S\ref{sc:photcal}), additional observations were obtained with the 3.5-m WIYN telescope at Kitt Peak National Observatory using the MiniMosaic camera. Its field of view of $9\farcm6\times9\farcm6$ is covered by 2 CCDs. The camera was used in $2\times2$ binned mode, which yields an effective scale of $0\farcs28/$pixel. We observed ten fields covering \ngal\ at three different epochs (one night per lunation for three consecutive months) using SDSS {\it gri} filters (Kitt Peak filter numbers k1017, k1018, k1019). The location of the fields is outlined in gray (blue in online edition) in Figure~\ref{fig:mos}. An additional four fields covering M67 were observed to derive accurate color transformations. Exposure times of 30\,s, 300\,s and 600\,s (hereafter, ``shallow'', ``medium'' and ``deep'') were chosen to bridge the magnitude gap between SDSS and our Gemini photometry.

\vfill

\section{Data Reduction and Photometry \label{sc:drphot}}

\subsection{Gemini \label{sc:drgem}}

We processed the raw images using the IRAF\footnote{IRAF is distributed by the National Optical Astronomy Observatory, which is operated by the Association of Universities for Research in Astronomy (AURA) under cooperative agreement with the National Science Foundation.} {\tt gemini} package. These routines perform overscan, bias and flat-field corrections that take into account the unique field of view of GMOS. Each CCDs was extracted to a separate FITS file, and the edges were trimmed by an additional 50 pixels.

\vspace{3pt}

Due to the crowded nature of the fields, we carried out point-spread function (PSF) photometry using the {\tt DAOPHOT} and {\tt ALLSTAR} programs \citep{stetson87,stetson93} on each image. Through visual inspection of the images using IRAF, we derived a starting value for the PSF FWHM of 5 pixels with a local sky annulus extending from 15 to 20 pixels. The task {\tt FIND} was used for an initial detection of objects above a 5$\sigma$ while the {\tt PHOT} task returned aperture photometry for these objects. Stars at or near the saturation limit and objects within $2-5\arcsec$ were identified and temporarily removed from the photometry files to ensure they were not used in the calculation of the PSF model. Saturation trails were masked in a similar manner. The {\tt PICK} task was used to select 100 stars from the cleaned aperture photometry list, which were visually examined to confirm that they were bright and isolated and to reject misidentified galaxies and stars with close companions. About 15-35 stars per chip remained after this examination, which were used by the {\tt PSF} task to calculate a PSF model for each image. Finally, {\tt ALLSTAR} was run to obtain preliminary PSF photometry for all sources.

\vspace{3pt}

We used {\tt DAOMATCH} and {\tt DAOMASTER} to calculate coordinate transformations between the images. We selected 7 or 8 images with the best seeing to create a master image in each band and chip. We performed photometry on each master frame as described above, but this time adopting a $3\sigma$ threshold. The total number of objects detected was $\sim4\times10^4$, $6\times10^4$ and $7\times10^4$ in {\it gri} respectively. Lastly, {\tt ALLFRAME} \citep{stetson94} was used to carry out fixed-position, simultaneous PSF photometry on all images.

\subsection{WIYN \label{sc:drwiyn}}

We applied an overscan, bias and flat-field correction on all images obtained at the WIYN telescope using the IRAF {\tt mscred} package. We performed PSF photometry on all images using the {\tt DAOPHOT} package as described in the previous section. We selected bright, isolated stars to create a PSF model for each image and {\tt ALLSTAR} was run to obtain PSF photometry.

\subsection{Photometric Calibration \label{sc:photcal}} 

Due to the significant difference in the magnitude range covered by the SDSS-DR7 photometric catalog \citep{abazajian09} and our Gemini images, it was not possible to obtain a direct calibration of the latter based on the former. Bright stars in the Gemini images were undetected by SDSS, while most bright SDSS stars were saturated in the Gemini fields. We bridged this magnitude gap by observing the \ngal\ fields with WIYN as described in \S\ref{sc:wiynobs} and generating a catalog of local standards.

\vspace{3pt}

All steps in our photometric calibration procedure are listed in Table~\ref{tab:photcal}. We describe the term being solved for, the source and target photometric catalogs, magnitude range of the stars being used, number of objects used in the final fit, and the systematic uncertainty to be propagated into our final Cepheid magnitudes. In the case of color terms, we evaluated the uncertainty at the extremes of the color range spanned by Cepheids ($\pm0.5$~mag relative to the pivot color used in our solutions). In all cases we used PSF photometry and parameters were determined through an iterative sigma clipping procedure. We visually inspected all objects being used in any step that tied two different telescopes/cameras to remove galaxies and blends. Some comments on the individual steps follow.

\begin{deluxetable}{llllllrrrr}
\tablecaption{Photometric calibration steps\label{tab:photcal}}
\tabletypesize{\scriptsize}
\tablewidth{\textwidth}
\tablehead{\colhead{Step} & \colhead{Reference} & \colhead{Target}    & \multicolumn{3}{c}{Mag.~range}                & \colhead{$N$}   & \multicolumn{3}{c}{$\sigma$ [mmag]}          \\
                          &                     &                     & \colhead{$g$} & \colhead{$r$} & \colhead{$i$} & \colhead{stars} & \colhead{$g$} & \colhead{$r$} & \colhead{$i$}}
\startdata
WIYN/MiMo color term  & SDSS-DR7 cat      & M67               & $13\mm 21$ & \multicolumn{2}{c}{$12\mm 20$} &               345 &  2 &  2 &  1 \\
WIYN/MiMo zeropoint   & SDSS-DR7 cat      & \ngl\ ``shallow'' & $14\mm 20$ & $13\mm 19$ & $14\mm 19$        &                79 &  4 &  6 &  9 \\
WIYN/MiMo transfer    & \ngl\ ``shallow'' & \ngl\ ``deep''    & \multicolumn{3}{c}{$16\mm 21$}              &                70 & 20 & 21 & 30 \\
Gem./GMOS-N color term& \multicolumn{5}{l}{Adopted from \citet{jorgensen09}}                                &              \nd  &  1 &  2 &  3 \\
Gem./GMOS-N zeropoint & \ngl\ ``deep''    & GMOS inner/outer  & \multicolumn{3}{c}{$21\mm 24$}              &               101 & 38 & 33 & 32 \\
\tableline
Total                 &                   &                   &            &            &                   &                   & 43 & 40 & 45  
\enddata
\tablecomments{Systematic uncertainties associated with color terms are evaluated at the extreme ranges of Cepheid colors. All quoted values are averages over different CCDs; actual values were propagated for the Cepheid photometry.}
\end{deluxetable}

\vspace{3pt}

We found small but well-detected color terms for the transformation of WIYN MiniMosaic magnitudes into the SDSS system; using $g\mm r$ as the target color, the values were $-0.038, -0.032, -0.037\pm0.003$ for $gri$, respectively. \hspace{2pt} These were \hspace{1pt} derived using \hspace{1pt} high {\it SNR} obser-

\ \par

\vspace{1.73in}

\noindent{vations of M67 and were fixed for the subsequent step (determination of zeropoints for the ``shallow'' \ngal\ fields). Table~\ref{tab:secstd} lists the magnitudes of these secondary standards, which may be useful to future observers. Due to the limited color range of the stars in common between WIYN and Gemini, and their noisier photometry (median $\sigma\ig 0.045$~mag), we adopted the color terms for GMOS-N derived by \citet{jorgensen09} and only solved for the zeropoints. We listed the mean uncertainties for this step in the Table, but propagated the actual values in our calculations. In summary, we estimate systematic zeropoint uncertainties of $\sim 45$~mmag for our Cepheid magnitudes.}

\vspace{3pt}

We carried out artificial star tests to characterize the completeness and crowding biases in the Gemini photometry. We divided the color-magnitude diagram into four quadrants and randomly selected 30 stars from each one to ensure that a broad range of stars were simulated. We added these 120 stars to the master frame with the {\tt DAOPHOT} task {\tt ADDSTAR}. We repeated this procedure 20 times to increase the statistical significance of our simulations. We performed photometry and matched the detected objects with the input artificial star lists, adopting a critical matching radius of 1.1~pix (equivalent to $3\sigma$). Table~\ref{tab:cfrac} lists the magnitudes at which we expect to detect 50\% and 80\% of the sources. We found no statistically significant photometric bias due to crowding at the magnitudes equivalent to 50\% completeness levels. Given the maser distance to \ngal\ and the fiducial P-L relations discussed in \S\ref{SC:SLOANPL}, we expect our Cepheid sample to be severely incomplete below $P\ig 10$ and $15$d for the outer and inner fields, respectively. 

\vspace{3pt}

Before discussing the identification of Cepheid variables in our data, we will address the issue of fiducial Cepheid P-L relations in the SDSS filters since these are used in our candidate selection process.

\section{Fiducial Cepheid P-L relations in SDSS filters \label{sc:sloanpl}}

\begin{deluxetable}{llrrrrrrl}
\tabletypesize{\scriptsize}
\tablecaption{Secondary standards \label{tab:secstd}}
\tablewidth{0.49\textwidth}
\tablehead{\colhead{RA} & \colhead{Dec.} & \multicolumn{3}{c}{Magnitudes} & \multicolumn{3}{c}{$\sigma$ [mmag]} & \colhead{Used}\\
\multicolumn{2}{c}{[deg, J2000]} & \colhead{$g$} & \colhead{$r$} & \colhead{$i$} & \colhead{$g$} & \colhead{$r$} & \colhead{$i$} & \colhead{in}}
\startdata
184.39152 & 47.24800 & 16.764 & 16.256 & 16.141 &   3 &   6 &   6 & SW \\
184.39327 & 47.25668 & 20.676 & 19.296 & 18.628 &   6 &   7 &   7 & SW \\
184.39842 & 47.22298 & 19.955 & 19.497 & 19.290 &  10 &  18 &  18 & SW \\
184.42695 & 47.24926 & 19.917 & 18.898 & 18.552 &   4 &   4 &   4 & SW \\
184.43053 & 47.22627 & 19.986 & 19.540 & 19.388 &   5 &   9 &   9 & SW
\enddata
\tablecomments{SW: used in SDSS-WIYN calibration; WG: used in WIYN-Gemini calibration. This table is available in its entirety in machine-readable form in the online version of the paper. A portion is shown here for guidance regarding its form and content.\vspace{-9pt}}
\end{deluxetable}

\begin{deluxetable}{lllllll}
\tablecaption{Photometric completeness limits\label{tab:cfrac}}
\tablewidth{0.49\textwidth}
\tablehead{
\colhead{Field} & \multicolumn{3}{c}{80\% completeness}         & \multicolumn{3}{c}{50\% completeness} \\
                & \colhead{$g$} & \colhead{$r$} & \colhead{$i$} & \colhead{$g$} & \colhead{$r$} & \colhead{$i$}}
\startdata
Inner           &  24.7         &  25.0         &  24.2         &  25.9         &  25.5         &  24.9 \\
Outer           &  25.2         &  25.1         &  24.5         &  26.1         &  25.7         &  25.3 
\enddata
\end{deluxetable}

\begin{figure}[htbp]
\begin{center}
\includegraphics[width=0.49\textwidth]{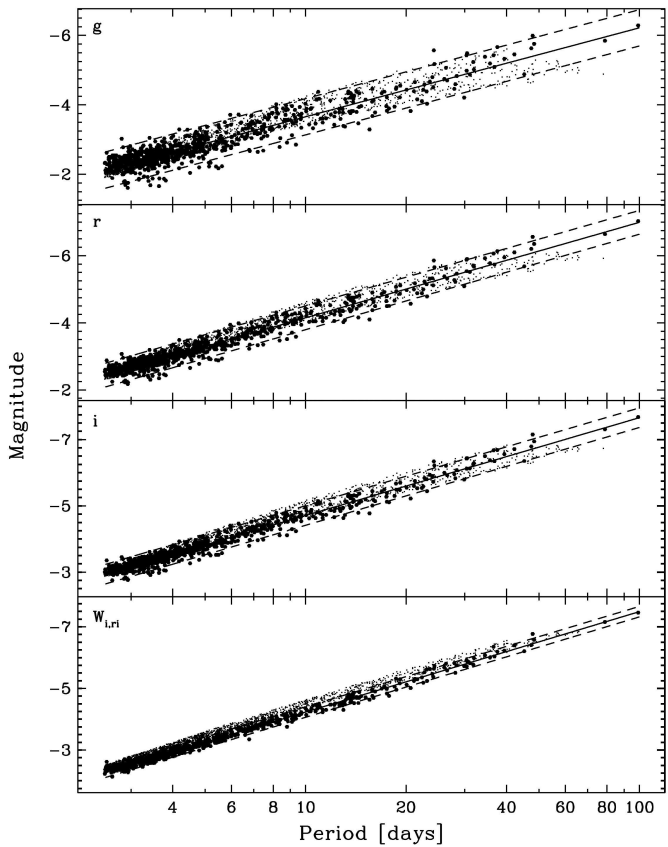}
\end{center}
%\vspace{-6pt}
\caption{Fiducial Cepheid P-L relations in the SDSS {\it gri} bands (top to bottom). Filled symbols denote LMC Cepheids, transformed from {\it VI} to {\it gri} using the procedure described in \S\ref{SC:SLOANPL}, while dots represent theoretical Cepheid magnitudes from \citet{dicriscienzo13}. The solid line represents the best fit to the LMC data while the dashed lines indicate the $\pm2\sigma$ width of the relations.\label{fig:plfid}}
\end{figure}

Despite its introduction nearly two decades ago, the SDSS filter set has rarely been used for Cepheid photometry. The two most notable uses are the massive surveys of M33 \citep{hartman06} and M31 \citep{kodric13,kodric14} using the MMT and the Pan-STARRS telescopes, respectively. Unfortunately, despite concerted efforts over the past decade \citep{ribas05,bonanos06,vilardell10} neither galaxy has a distance estimate as robust as that for the LMC by \citet{pietrzynski13}: $D\ig 49.97\pm2\%$~kpc (equivalent to $\mu_0\ig 18.493\pm0.048$~mag). Furthermore, given the apparent LMC-like metallicity prevalent throughout most of the disk of \ngal\ \citep{bresolin11}, this Milky Way satellite should provide the most appropriate sample of Cepheids from which to obtain a fiducial P-L relation for our analysis.

\vspace{3pt}

Motivated by the above, and in a manner similar to previous work by \citet{ngeow07}, we generated semi-empirical Cepheid P-L relations in the SDSS {\it gri} filters based on {\it VI} photometry for $\!>\!750$~LMC variables with $2.5\!<\!P\!<\!100$d compiled by \citet{macri14}. This dataset consists mainly of OGLE photometry \citet{soszynski08,ulaczyk13} supplemented by literature measurements for additional long-period objects \citep{martin79,freedman85,barnes99,tanvir99,sebo02,ngeow06}. We derived photometric transformations appropriate for Cepheids using synthetic magnitudes for stars with $\log g \leq 1$ based on the \citet{castelli03} models, kindly provided by F.~Castelli\footnote{http://wwwuser.oats.inaf.it/castelli/colors.html}. We fit cubic-order polynomials to stars with $V\mm I\!<\!1.5$ and obtained transformations with ${\rm \it rms}\!<\!0.01$~mag. Using the previously-discussed distance modulus for the LMC, this procedure yielded the following P-L relations in the SDSS {\it gri} filters:
\begin{eqnarray}
g &=& -3.657(50) - 2.560(34) (\log P\mm 1) \ \ \ \ \sigma\ig 0.261 \\
r &=& -4.148(49) - 2.845(23) (\log P\mm 1) \ \ \ \ \sigma\ig 0.177 \\
i &=& -4.275(48) - 2.952(19) (\log P\mm 1) \ \ \ \ \sigma\ig 0.148 
\end{eqnarray}

\noindent{where the zeropoint uncertainties include the term associated with the distance modulus. We then calculated an independent set of P-L relations based on the theoretical Cepheid magnitudes in SDSS filters computed by \citet{dicriscienzo13}. We restricted the dataset to $2.5\!<\!P\!<\!40$d due to the incomplete filling of the instability strip beyond the upper period limit, which arises as a consequence of the upper mass limit considered in the models. We obtained:}
\begin{eqnarray} 
g &=& -3.738(07) - 2.615(18) (\log P\mm 1) \ \ \ \ \sigma\ig 0.214 \label{eqn:plg}\\
r &=& -4.241(05) - 2.882(13) (\log P\mm 1) \ \ \ \ \sigma\ig 0.161 \\
i &=& -4.402(04) - 2.987(12) (\log P\mm 1) \ \ \ \ \sigma\ig 0.139 \label{eqn:pli}
\end{eqnarray}

\noindent{which are in excellent agreement in terms of the slopes with the previous set of relations; both sets are shown in Figure~\ref{fig:plfid}. We used each set of PLs to derive relations between the residuals of a given Cepheid in two bands, which we will use in our candidate selection process below. We found:}

\begin{eqnarray}
\Delta r & = & 0.752\ \Delta g \ \ \sigma\ig 0.028\\
\Delta i & = & 0.650\ \Delta g \ \ \sigma\ig 0.038\\
\Delta i & = & 0.864\ \Delta r \ \ \sigma\ig 0.015,
\end{eqnarray}

\noindent{where the dispersions were calculated using the LMC data. We also calculated the $1\sigma$-equivalent ranges spanned by the variables along the color-color relations, which were 0.27, 0.25 and 0.21~mag, respectively.}

\section{Identification of Cepheid Variables \label{sc:idceph}}

We used the TRIAL program (kindly provided by P.~Stetson) \hfill to \hfill identify \hfill variable \hfill objects \hfill by \hfill calculating

\begin{deluxetable}{lr}
\tablecaption{Cepheid selection steps\label{tab:sel}}
\tablewidth{0pc}
\tablehead{\colhead{Step} & \colhead{Number}}
\startdata
$L_r \geq 0.75$   & 4419 \\
$N_r, N_i \!>\! 75\%$ & \nvar \\
%g photometry             
$A_i \geq 0.1$~mag& 2530 \\
non-aliased $P$   &  959 \\
``ABC'' grades    &  408 \\
pass visual insp. &  309
\enddata
\end{deluxetable}

\ \par

\noindent{the modified Welch-Stetson variability index $L$ \citep{stetson96} in the $r$-band data, setting $L_r\ig 0.75$ as the variability threshold and only considering objects with valid photometry in $\geq 75$\% of the $r$ \& $i$ images. There were \nvar\ objects that met these criteria; of these, \gfrac\% also had valid $g$ photometry. We only expected a small fraction of the variables to be Cepheids, with the majority likely being irregular or semi-periodic RGB/AGB pulsators. We selected Cepheid candidates following the steps outlined below; the number of objects rejected at each stage are summarized in Table~\ref{tab:sel}.}

\begin{figure}[htbp]
\begin{center}
\includegraphics[width=0.49\textwidth]{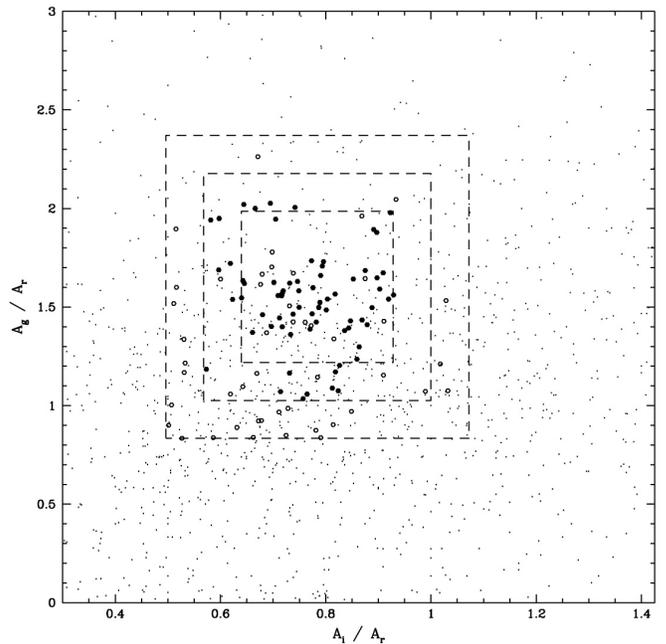}
\end{center}
\vspace{-6pt}
\caption{Amplitude ratios derived from the best-fit Cepheid template light curves for all variables with $L_r \geq 0.75$. Dashed lines indicate the various regions used to grade variables (A, B, C or F) based on the amplitude ratios spanned by LMC Cepheids. Filled symbols denote objects listed in Table~\ref{tab:ceph}; open symbols represent objects listed in Table~\ref{tab:vars}, and small dots represent variables rejected at any step of the selection process.\label{fig:amprat}}
\end{figure}

\renewcommand{\theenumi}{\alph{enumi}}
\begin{enumerate}
\item We ran the Cepheid template-fitting program developed by \citet{yoachim09} on the $(g)ri$ light curves, using 100 initial trial periods spanning 7 to 124 days (spaced every 0.0125 dex in $\log P$). The lower limit was set by our sparse observational sampling and estimated completeness limit (described in \S\ref{sc:photcal}) while the upper limit was set to search for ultra-long period Cepheids. We selected the best-fit period corresponding to the lowest value of $\chi^2$ returned by the template-fitting program. We derived flux-weighted mean magnitudes by numerical integration of the best-fit template light curves, and calculated the light curve semi-amplitudes as half of the difference between the faintest and brightest points in the template. The uncertainties in both of these parameters were estimated by evaluating $\chi^2$ over a grid of values while keeping the period fixed to the best-fit value.

\item We discarded objects with $i$-band semi-amplitudes below 0.1~mag to remove blended objects and low-amplitude semi-regular variables. We generated histograms of the best-fit periods for the remaining variables to identify any possible aliasing due to the sparse nature of our observations. Using a bin size of $\Delta\log P\ig 10^{-3}$, we found that $\sim 70\%$ of the bins were empty and $\sim 25\%$ of the bins had only one variable. We flagged any bin with more than 4 variables as a possibly aliased period and reran the previous step excluding those periods from consideration. We identified any remaining aliased periods after the second iteration and removed those objects from further consideration.

\item We carried out the template-fitting procedure described in (a) on the $BVI$ and $VI$ light curves of all fundamental-mode LMC Cepheids from OGLE-II \citep{udalski99} and OGLE-III \citep{soszynski08,ulaczyk13}, respectively, except that we kept the periods fixed to the published values. We transformed the resulting best-fit templates into the $gri$ system using the previously-mentioned models by \citet{castelli03} and calculated the light curve amplitude ratios exhibited by Cepheids in these bands. We found $A_g/A_r\ig 1.610\pm0.062$ and $A_i/A_r\ig 0.781\pm0.024$. We then classified the remaining variables in \ngal\ according to their amplitude ratios; objects within $6\sigma$ of the LMC values were given a grade of ``A'', those at $6-9\sigma$ ``B'', those at $9-12\sigma$ ``C'', and the rest ``F''. Variables without valid $g$ photometry were classified solely based on their $i$-to-$r$ amplitude ratio. Figure~\ref{fig:amprat} shows the result of this step.

\end{enumerate}

\begin{figure}[htbp]
\begin{center}
\includegraphics[width=0.49\textwidth]{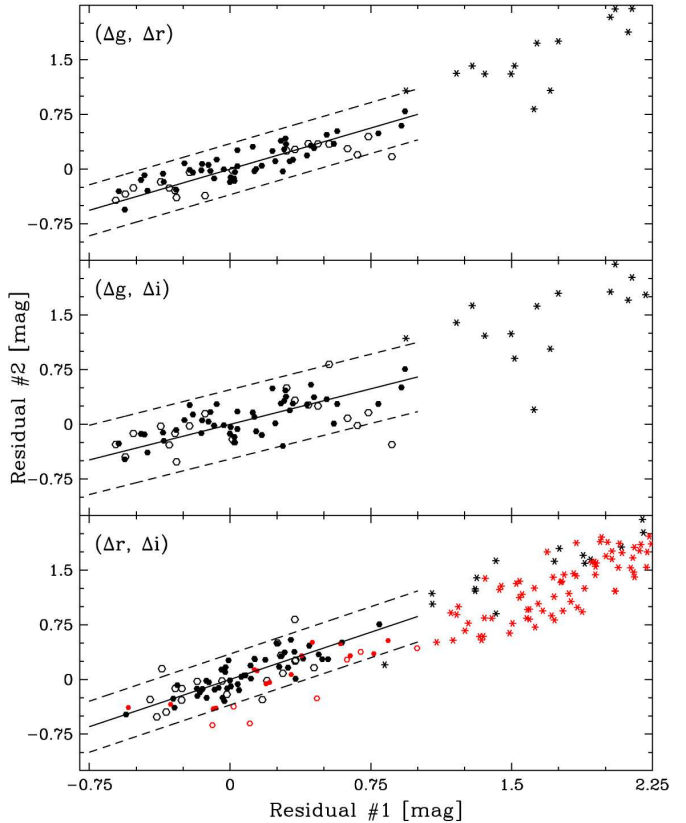}
\end{center}
\caption{Correlation of P-L residuals for all objects listed in Tables~\ref{tab:ceph} \&\ref{tab:vars}, relative to the best-fit P-L relations for objects with ``A''-grade amplitude ratios and $15 < P < 100$d. Filled symbols denote Cepheids with ``A'' grade in amplitude ratios and PL residuals while open symbols denote Cepheids with ``B'' grade in at least one category. Starred symbols represent objects listed in Table~\ref{tab:vars}. Red symbols are used for object with only $r$ and $i$ photometry.\label{fig:colcol}}
\vspace{3.25pt}
\end{figure}

\begin{enumerate}
\addtocounter{enumi}{3}
\item We selected variables with a grade of ``A'' from the preceding step, $gri$ photometry and $15\!<\!P\!<\!100$d as our reference subsets (to avoid incompleteness bias at the short end and possible non-linearities at the long end) and fit the P-L relations listed in Equations $4\mm 6$. We calculated the residuals of all variables in all bands relative to the best-fit relations and fit them using the relations listed in Equations $7\mm 9$. We flagged (with a grade of ``C'') and removed from further fitting any object with a residual in any band exceeding 1~mag in absolute value, as these are likely either badly blended Cepheids (on the bright side) or heavily reddened Cepheids/Pop II variables (on the faint side). We flagged (with a grade of ``B'') and removed from further fitting any objects lying beyond $6\sigma$ of the dispersions determined in Equations $7\mm 9$ {\it and} with residuals greater than $2.5\sigma$ based on the observed dispersion for \ngal\ Cepheids. We only flagged and removed one object per band on each iteration and continued until convergence. Figure~\ref{fig:colcol} shows the result of this step. 

\item Finally, we inspected the master images at the location of each variable to ensure that all candidates were well-resolved and isolated point sources, located at least $0\farcs5$ away from chip edges.
\end{enumerate}

The final Cepheid sample contains \nceph\ objects (listed in Table~\ref{tab:ceph}) that received a grade of ``A'' or ``B'' in steps (c) and (d). Variables with a grade of ``C'' in either step are listed in Table~\ref{tab:vars}; these \nvars\ objects are probably blends, highly reddened Cepheids, or Population II pulsators. The locations of both sets of objects within the Gemini fields are shown in Figures~\ref{fig:gmosout} \& \ref{fig:gmosinn}, while individual finding charts can be found in Figures~\ref{fig:fcha}-\ref{fig:fchg}. Representative light curves are plotted in Figure~\ref{fig:lcs} and all light curve data is presented in Table~\ref{tab:lcdata}.

\section{Results\label{sc:res}}

We calculated the Cepheid detection efficiency and robustness of the derived periods by comparing our sample with that of \citet{macri06} over the areas in common (see Figure~\ref{fig:mos}). There are 246 Cepheids from that study with $4\!<\!P\!<\!45$~d located within our fields. As expected from the artificial star tests described in \S\ref{sc:photcal}, our ability to detect significant variability ($L_r \geq 0.75$) was very low (9\%) for $P<7$~d Cepheids, increasing to 42\% and 56\% for $7<P<15$~d and $P>15$~d, respectively. Focusing on the last group, 53\% of the detected variables were ultimately rejected because of aliased periods or very low pulsation amplitudes (rejection criterion ``b'' in \S5), 41\% were classified as Cepheids, and 6\% classified as ``variables'' (highly reddened/blended Cepheids or Pop.~II pulsators). The periods we derived for the objects classified as Cepheids were very robust, with $\langle\Delta\log P\rangle\ig -0.005\pm0.010$ relative to their HST-based values. We also compared our results with those of \citet{fausnaugh14}. We detected significant variability for 79\% of their Cepheids located within our fields and classified 74\% of these as Cepheids, 4\% as lower-quality variables, and rejected the remaining 22\%. A comparison of the periods for Cepheids in common again revealed excellent agreement for all but one object, with $\langle\Delta\log P\rangle\ig -0.0007\pm0.0001$. After taking into account objects present in the two aforementioned studies, our survey contributes an additional 57 Cepheids \& 205 variables.

\vspace{3pt}

We present the P-L relations for Cepheids and variables in our sample in Figure~\ref{fig:pls}. The Cepheid relations become incomplete at $P\sim15$d, as expected from the artificial star tests and the detection efficiency discussed above. We fit the P-L relations listed in Eqns.~\ref{eqn:plg}-\ref{eqn:pli} to the 40 Cepheids in Table~\ref{tab:ceph} with $gri$ data and $15\!<\!P\!<\!100$d and obtained apparent distance moduli of $\mu_g\ig 29.29\pm0.06\rm{(r)}\pm0.04\rm{(s)}$, $\mu_r\ig 29.24\pm0.05\rm{(r)}\pm0.04\rm{(s)}$ and $\mu_i\ig 29.24\pm0.05\rm{(r)}\pm0.05\rm{(s)}$~mag (where {\it r} and {\it s} are used to denote random and systematic uncertainties, respectively). We adopted the extinction law of \citet{fitzpatrick99} with $R_V\ig 3.1$ and solved for the best-fit values of true distance modulus and reddening. Given the rather large uncertainties in the individual distance moduli and the short wavelength baseline provided by the filters we used, there is a large covariance between these two parameters. Nevertheless, we find $\mu_0\ig 29.18\pm0.23$~mag and $E(B\!-\!V)\ig 0.03\pm0.08$~mag, which are consistent at the $1\sigma$ level with the maser-based distance modulus of $\mu_0\ig 29.404\pm0.065$~mag \citep{humphreys13} and the foreground Galactic reddening towards \ngal\ of $E(B\!-\!V)\ig 0.014$~mag \citep{schlafly11}. Given the very shallow abundance gradient in \ngal\ \citep{bresolin11}, the Cepheids in our sample lie in areas of the disk that span a narrow range of LMC-like metallicities ($\langle [O/H] \rangle=8.34\pm0.07$~dex). We are therefore unable to provide any constraints on the ``metallicity effect''at these wavelengths \citep[for a recent study of this issue, see][]{fausnaugh14}.

\vspace{3pt}

Figure~\ref{fig:pls} also shows the expected P-L relations for Population II Cepheids in $r$ and $i$, which match fairly well the distribution of periods and magnitudes of the variables listed in Table~\ref{tab:vars}. The slopes of those relations were fixed to the values derived by \citet[][Table~3, entries labeled ``PLC'', which stands for clipped P-L relation]{kodric13} and the zeropoints were obtained by shifting the best-fit mean magnitudes of our observed P-L relations for ``classical'' (i.e., Population I) Cepheids at $P\ig 80$d by +1.91 mag. This average offset was derived by calculat-

\newcounter{subfigure}
\setcounter{subfigure}{1}
\renewcommand{\thefigure}{\arabic{figure}\alph{subfigure}}

\begin{figure*}[htbp]
\begin{center}
\includegraphics[width=\textwidth]{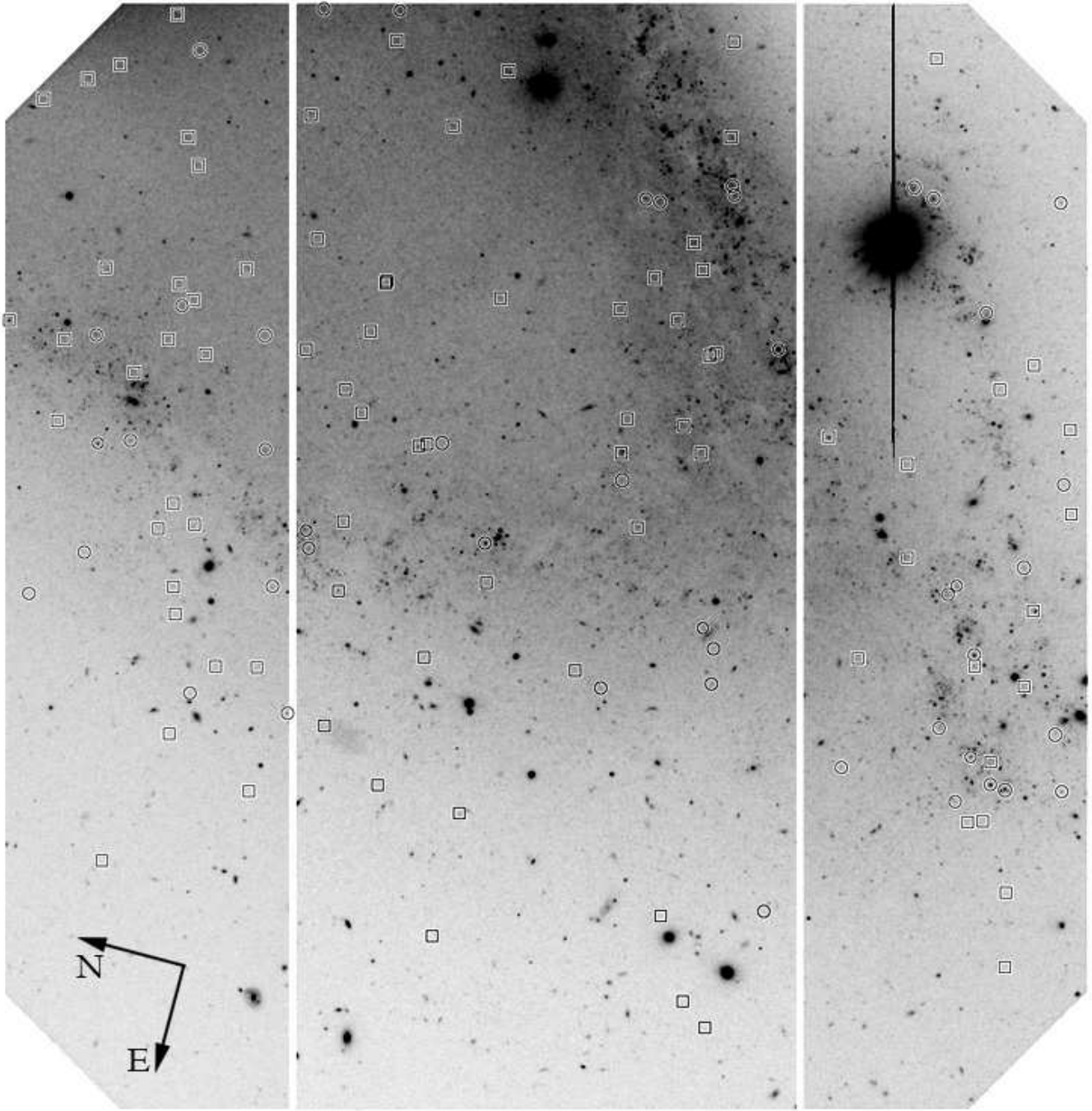}
\end{center}
\caption{$r$-band image of the Gemini outer field in \ngal. The locations of Cepheids (listed in Table~\ref{tab:ceph}) and variables (listed in Table~\ref{tab:vars}) are indicated by circles and squares, respectively. The image is $5\farcm5$ on a side.\label{fig:gmosout}}
\end{figure*}

\addtocounter{figure}{-1}
\addtocounter{subfigure}{1}

\begin{figure*}[htbp]
\begin{center}
\includegraphics[width=\textwidth]{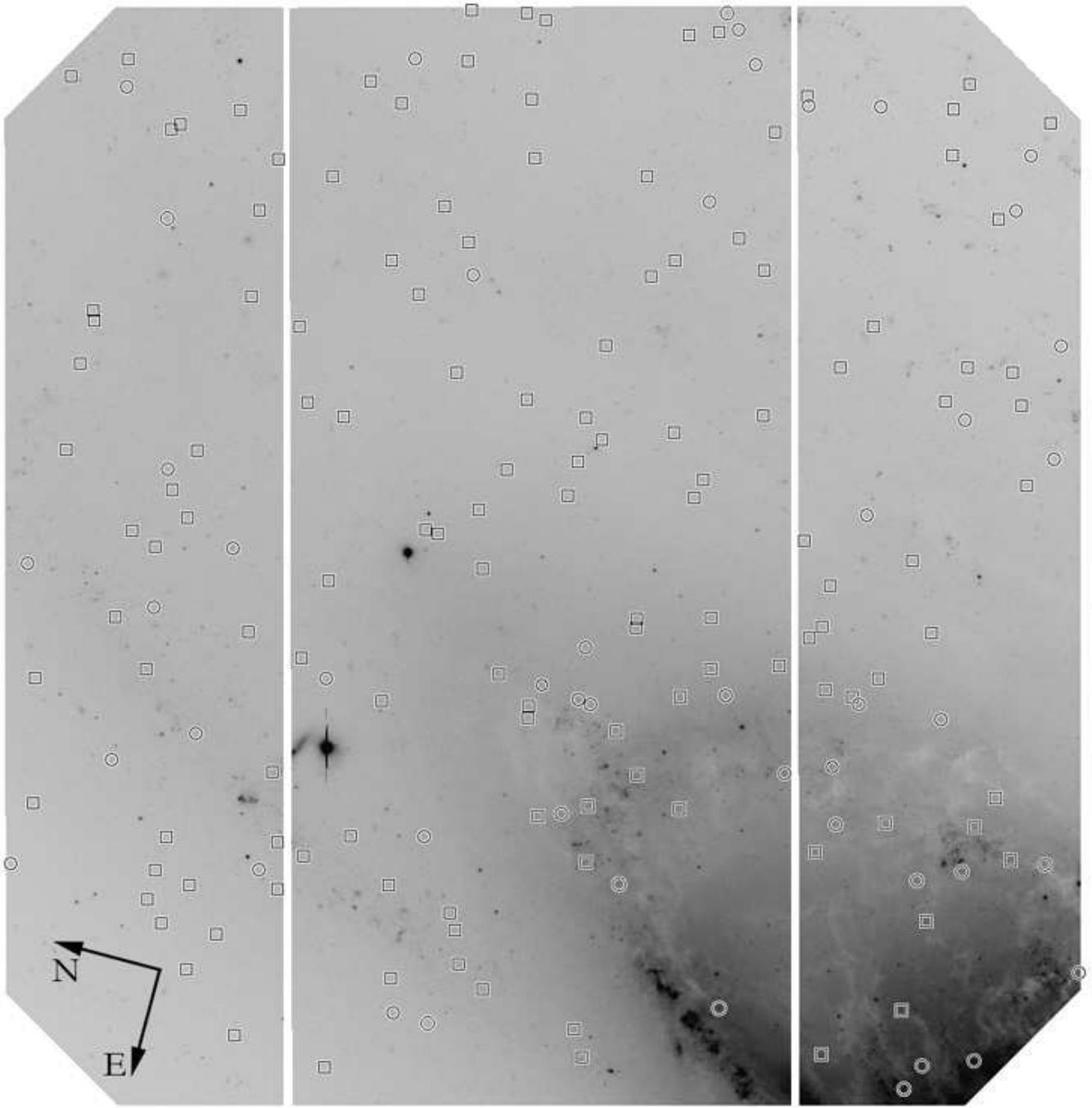}
\end{center}
\caption{Same as ~\ref{fig:gmosout}, but for the Gemini inner field in \ngal.\label{fig:gmosinn}}
\end{figure*}

\setcounter{subfigure}{1}

\begin{figure*}[htbp]
\begin{center}
\includegraphics[height=\textheight]{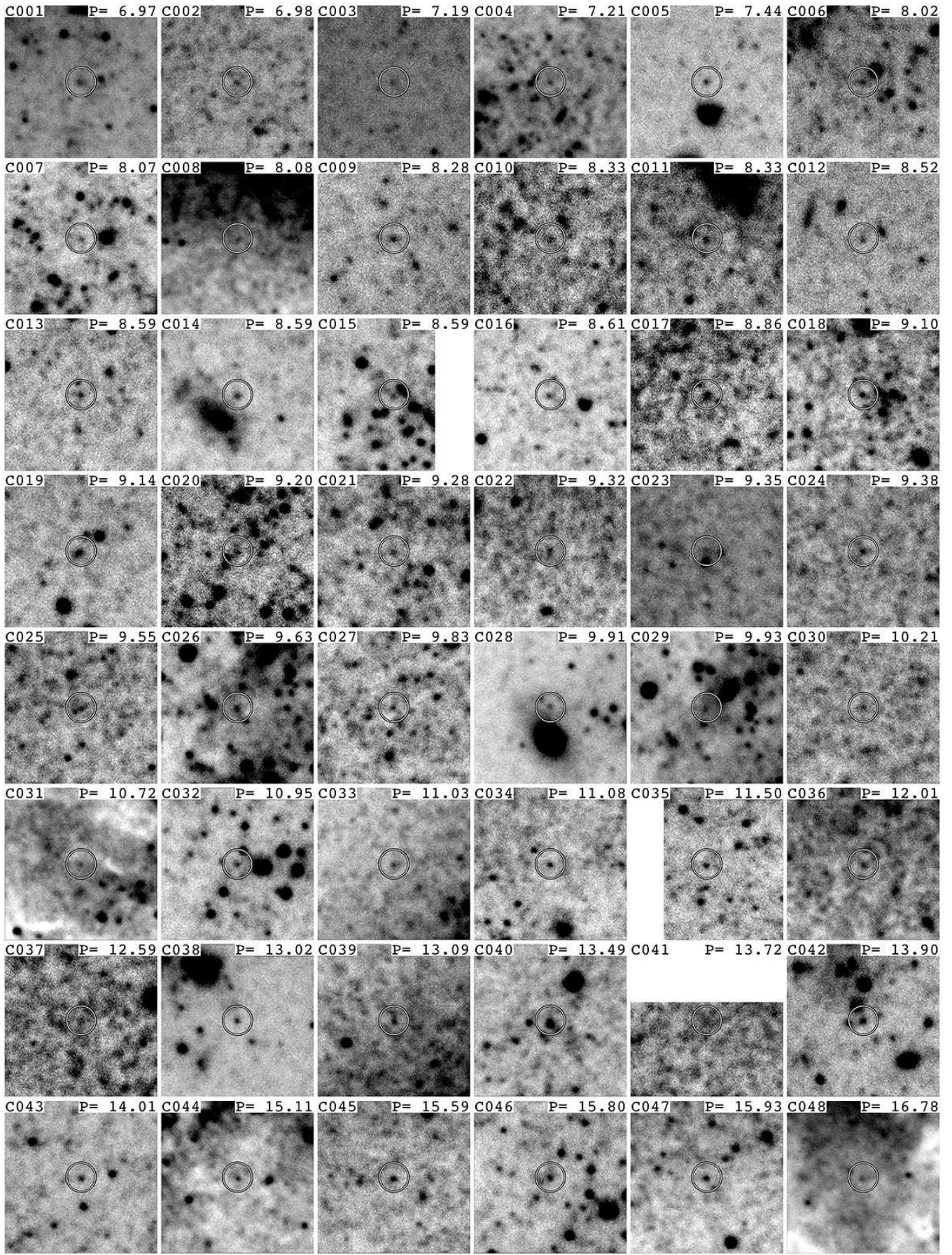}
\end{center}
\caption{Individual finding charts for the Cepheids and variables discovered in this work, listed in Tables~\ref{tab:ceph} and \ref{tab:vars}. Each panel is $14\farcs2$ on a side. \label{fig:fcha}}
\end{figure*}

\addtocounter{figure}{-1}
\addtocounter{subfigure}{1}

\begin{figure*}[htbp]
\begin{center}
\includegraphics[height=\textheight]{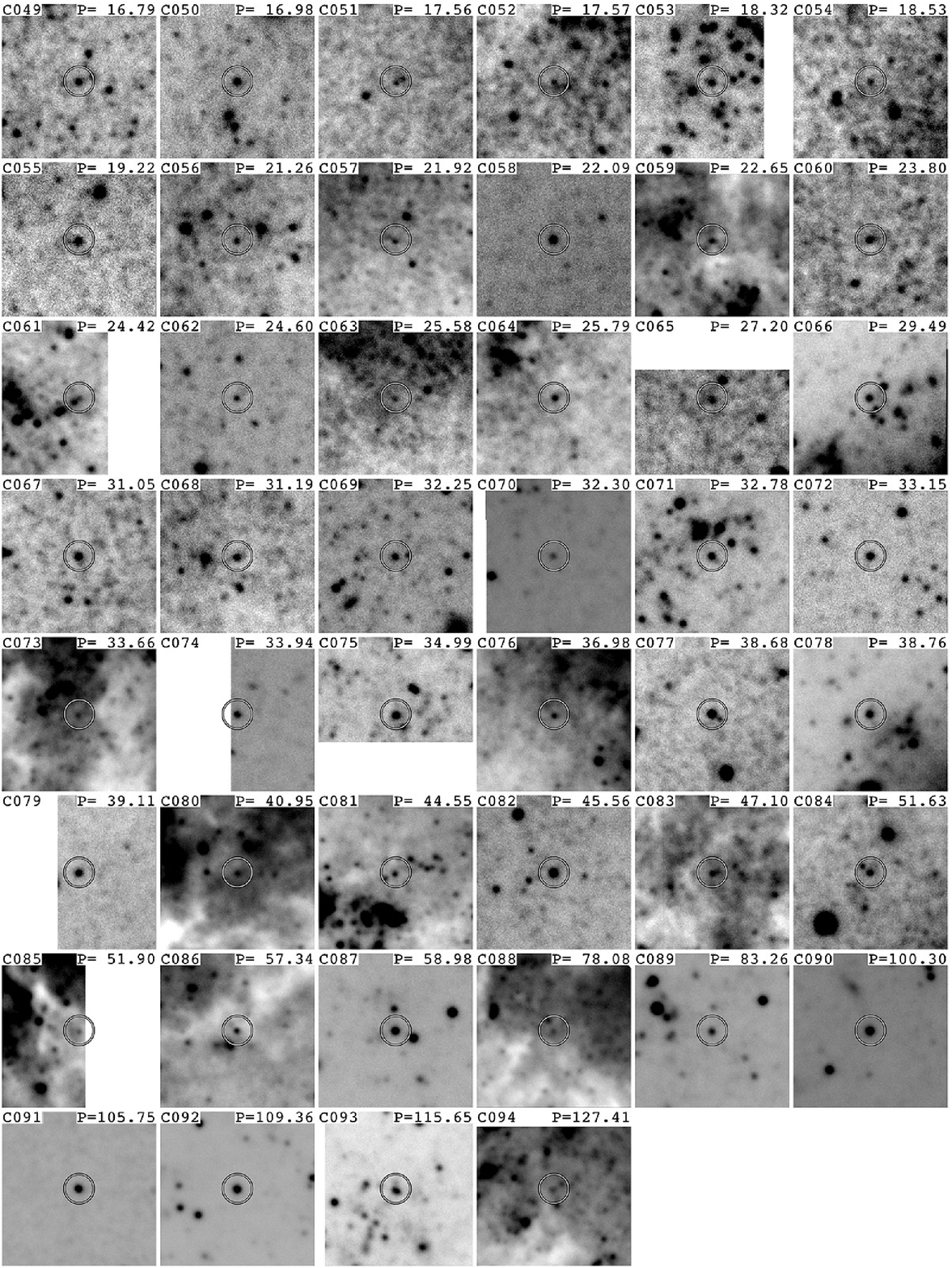}
\end{center}
\caption{continued. \label{fig:fchb}}
\end{figure*}

\addtocounter{figure}{-1}
\addtocounter{subfigure}{1}

\begin{figure*}[htbp]
\begin{center}
\includegraphics[height=\textheight]{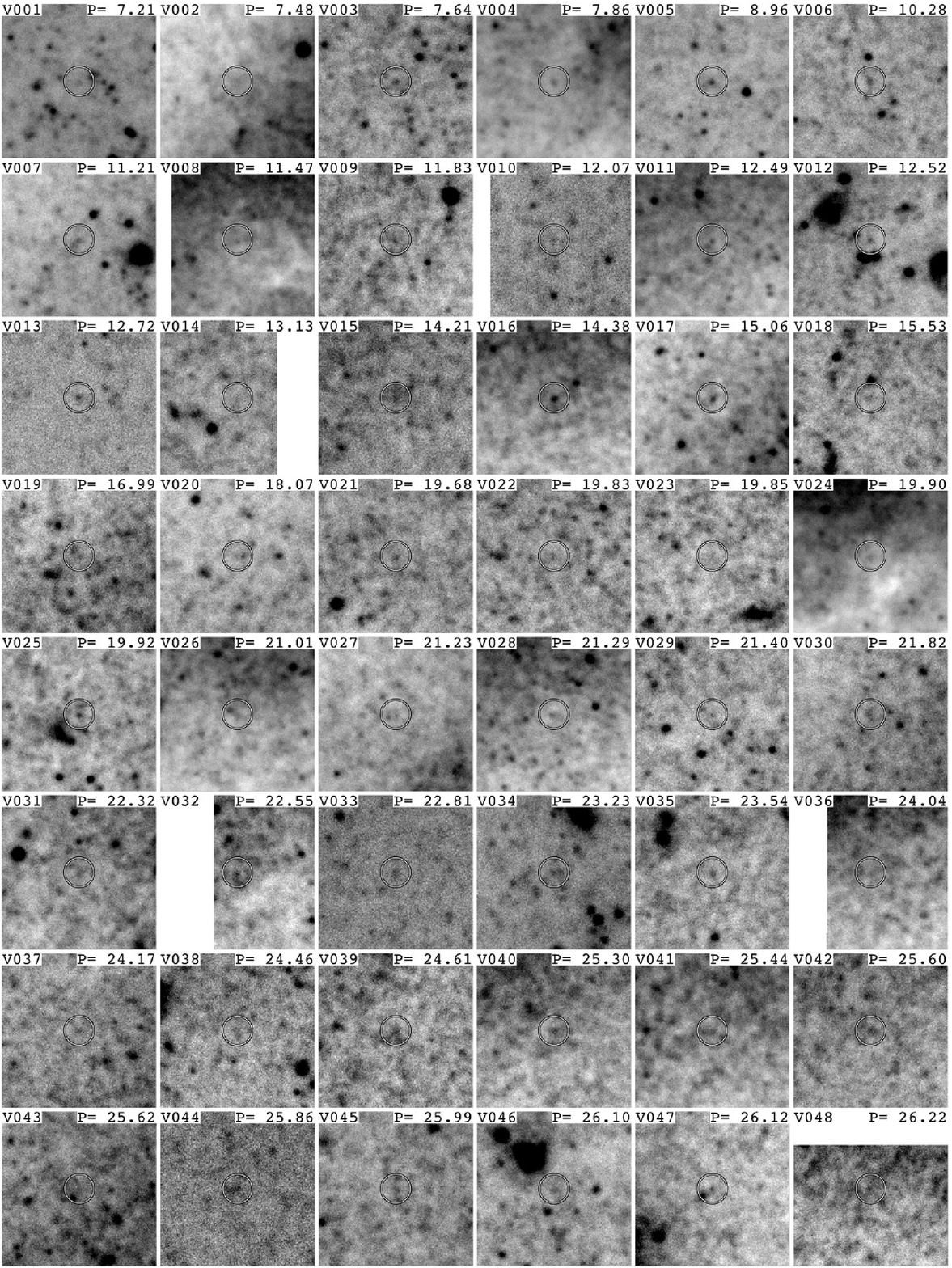}
\end{center}
\caption{continued. \label{fig:fchc}}
\end{figure*}

\addtocounter{figure}{-1}
\addtocounter{subfigure}{1}

\begin{figure*}[htbp]
\begin{center}
\includegraphics[height=\textheight]{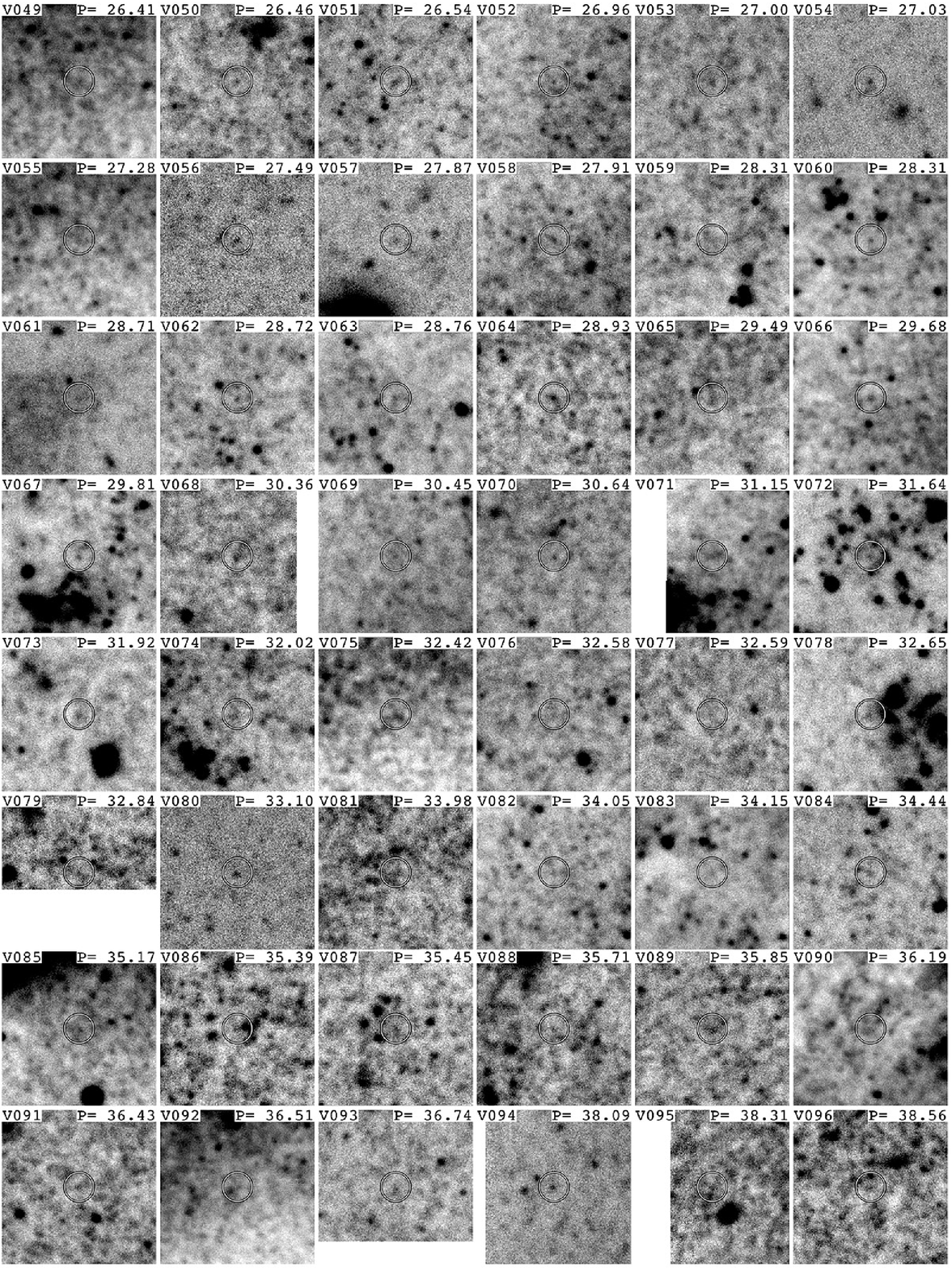}
\end{center}
\caption{continued. \label{fig:fchd}}
\end{figure*}

\addtocounter{figure}{-1}
\addtocounter{subfigure}{1}

\begin{figure*}[htbp]
\begin{center}
\includegraphics[height=\textheight]{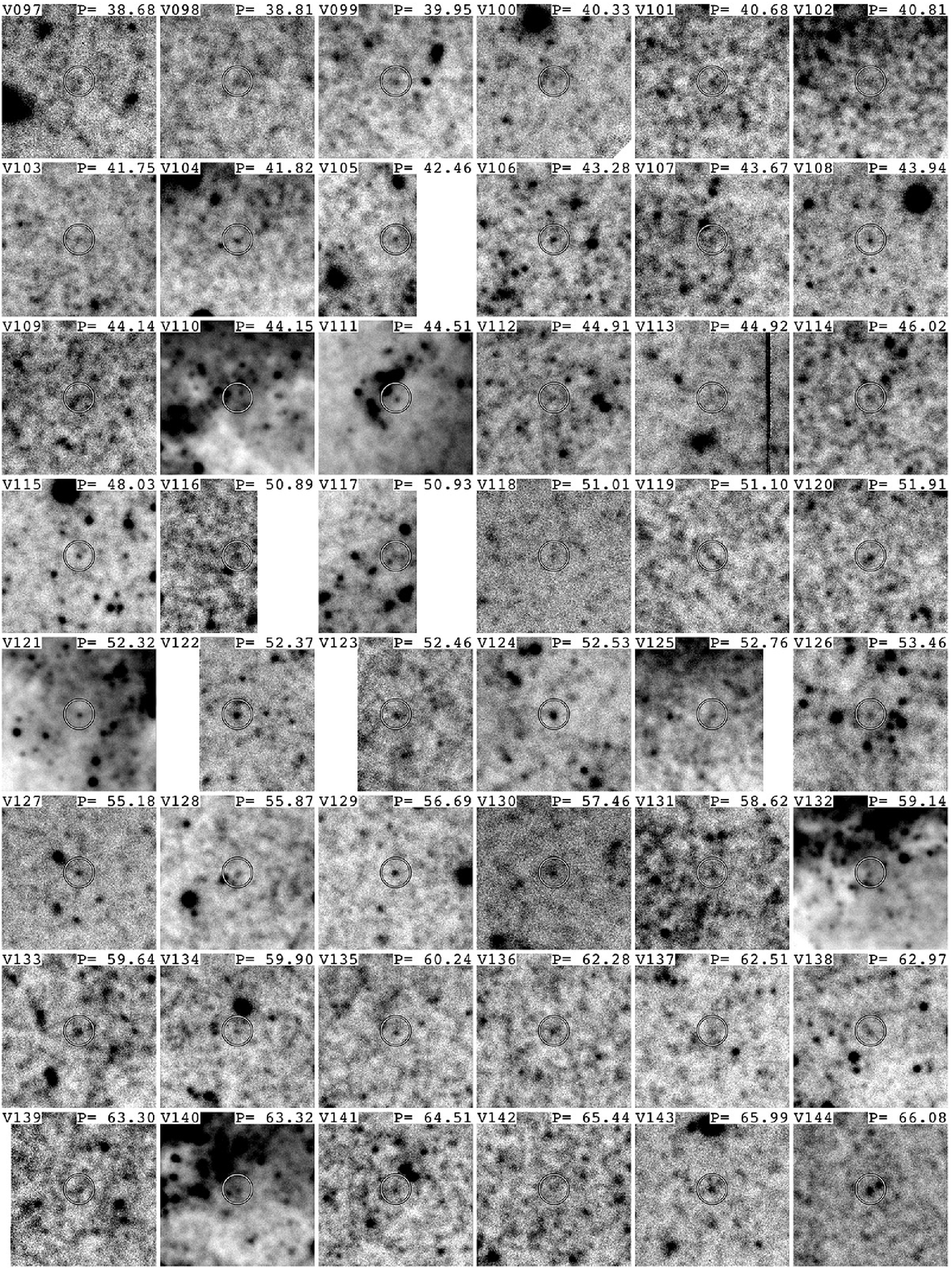}
\end{center}
\caption{continued. \label{fig:fche}}
\end{figure*}

\addtocounter{figure}{-1}
\addtocounter{subfigure}{1}

\begin{figure*}[htbp]
\begin{center}
\includegraphics[height=\textheight]{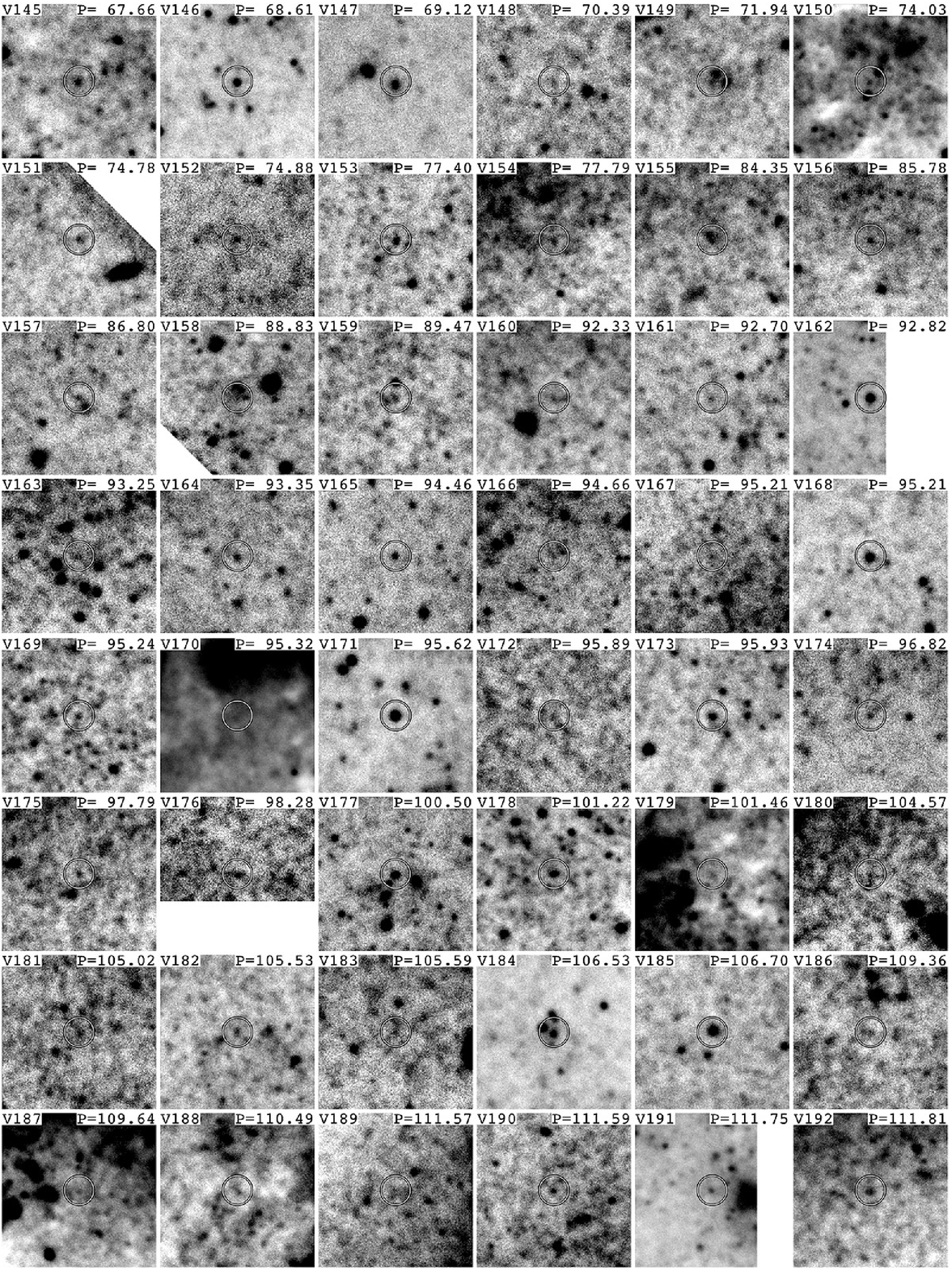}
\end{center}
\caption{continued. \label{fig:fchf}}
\end{figure*}

\addtocounter{figure}{-1}
\addtocounter{subfigure}{1}

\begin{figure*}[htbp]
\begin{center}
\includegraphics[height=\textheight]{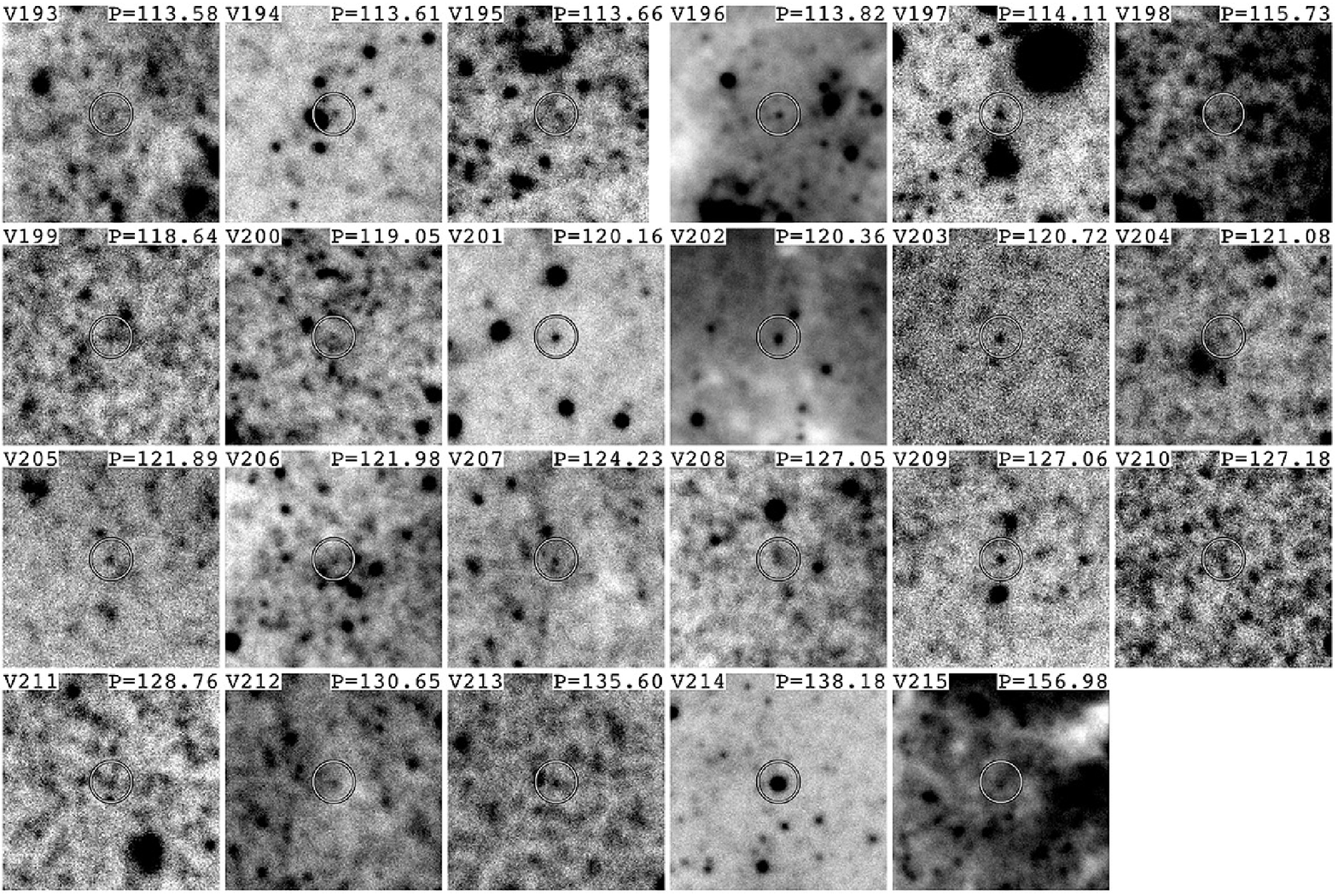}
\end{center}
\caption{continued. \label{fig:fchg}}
\end{figure*}

\renewcommand{\thefigure}{\arabic{figure}}

\begin{figure*}[htbp]
\begin{center}
\includegraphics[width=\textwidth]{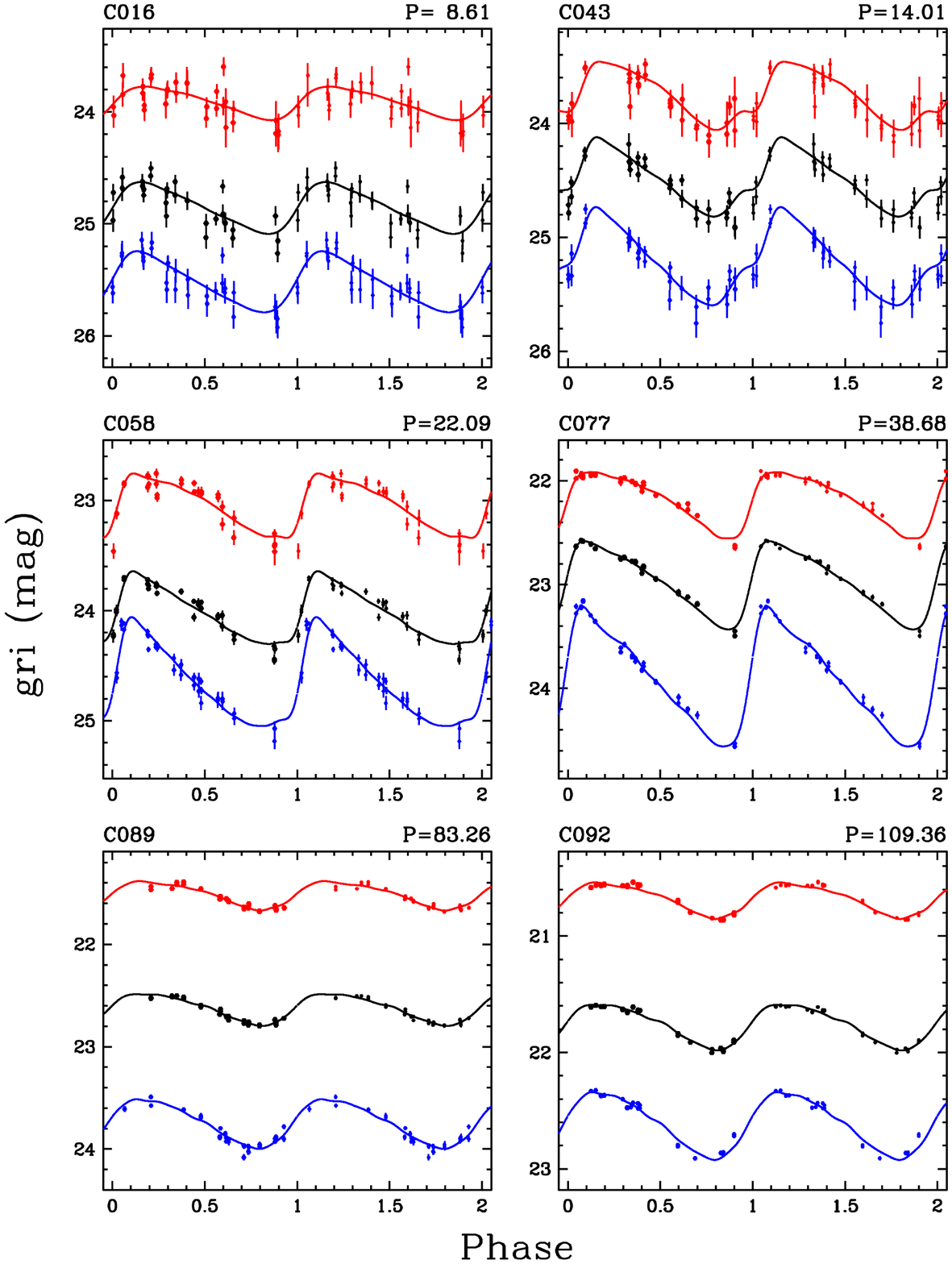}
\end{center}
\vspace{-18pt}
\caption{Representative Cepheid light curves. Filled symbols represent the Gemini photometry while the solid lines are the best-fit templates from \citet{yoachim09}. Offsets were added to the {\it gi} magnitudes and templates for clarity. \label{fig:lcs}}
\end{figure*}

\begin{figure*}[htbp]
\begin{center}
\includegraphics[width=0.9\textwidth]{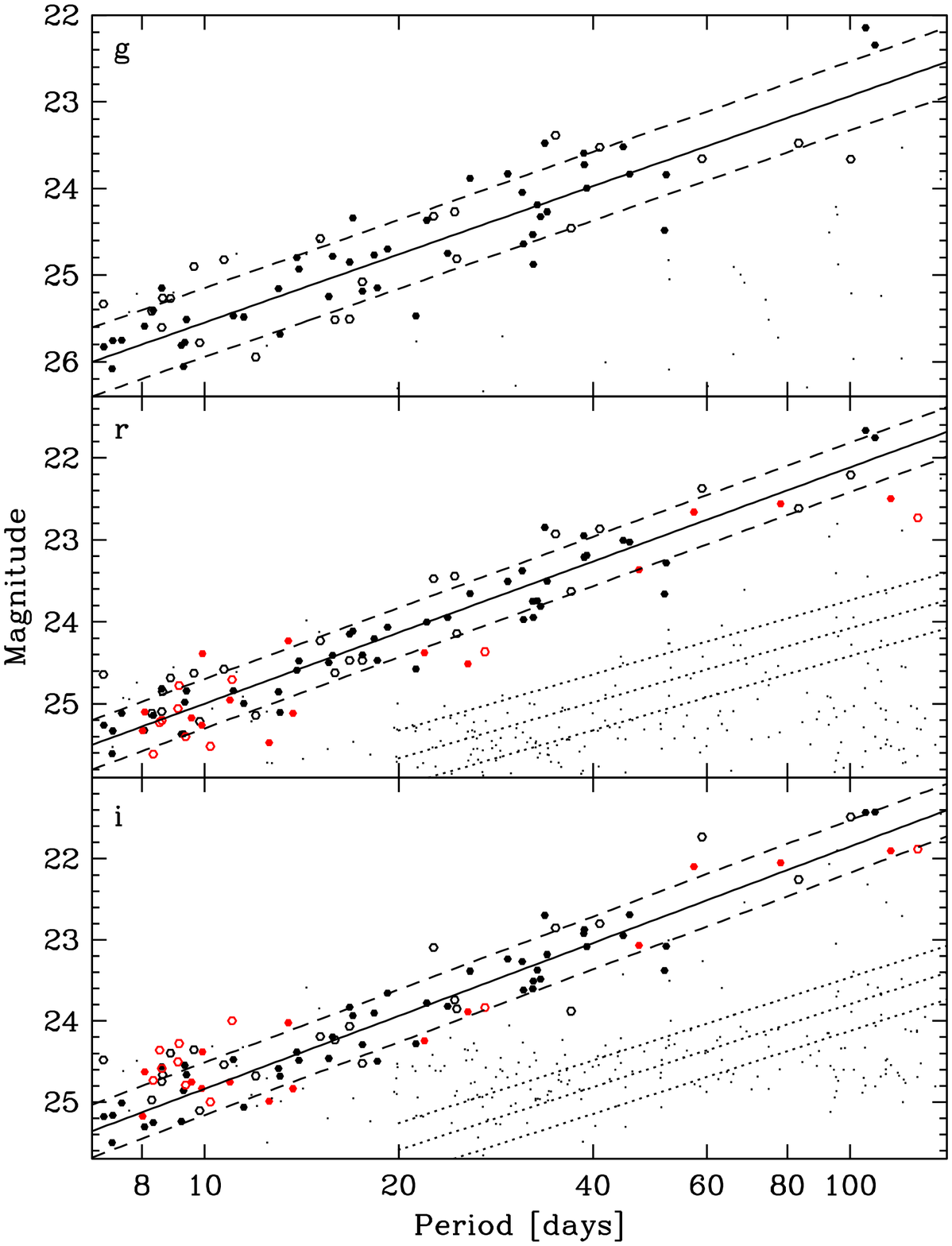}
\end{center}
\vspace{-18pt}
\caption{Period-Luminosity relations in {\it gri} (top to bottom) for Cepheids and other variables in \ngal. Filled symbols denote Cepheids with ``A'' grade in amplitude ratios and PL residuals while open symbols denote Cepheids with ``B'' grade in at least one category. Red symbols are used for Cepheids with only $r$ and $i$ photometry. Small dots represent objects listed in Table~\ref{tab:vars}. The uncertainties in mean magnitude and period are comparable to the size of the symbols. The slopes of the Cepheid P-L relations (solid lines) were fixed to the values derived from the theoretical Cepheid magnitudes of \citet{dicriscienzo13} as described in \S\ref{SC:SLOANPL}; the dashed lines indicate the $\pm2\sigma$ dispersion of the fits. The dotted lines in the $r$ and $i$ panels represent the Pop~II P-L relations of \citet{kodric13} shifted to the distance modulus of \ngal\ as described in \S\ref{SC:RES}.\label{fig:pls}}
\end{figure*}

\clearpage

\begin{deluxetable}{lllrrrrrrrrrrccl}
\tablecaption{\ngal\ Cepheids\label{tab:ceph}}
\tabletypesize{\tiny}
\tablewidth{\textwidth}
\tablehead{
\colhead{ID} & \colhead{RA} & \colhead {Dec} & \colhead{P}   & \multicolumn{6}{c}{Mean magnitudes} & \multicolumn{3}{c}{Lightcurve ampl.} & \multicolumn{2}{c}{Qual.} & \colhead{Cross}\\
\colhead{}   & \multicolumn{2}{c}{(J2000)}   & \colhead{}    & \colhead{$r$} & \colhead{$i$} & \colhead{$g$}  & \colhead{$\sigma_r$} & \colhead{$\sigma_i$} & \colhead{$\sigma_g$} & \colhead{$r$} & \colhead{$i$} & \colhead{$g$} & \multicolumn{2}{c}{flag} & \colhead{ID}   \\
\colhead{}   & \multicolumn{2}{c}{[deg]}     & \colhead{[d]} & \multicolumn{3}{c}{[mag]} & \multicolumn{3}{c}{[mmag]} & \multicolumn{3}{c}{[mmag]} & \colhead{A} & \colhead{R} & }
\startdata
C001 & 184.84502 & 47.24399 &  6.975 & 24.642 & 24.481 & 25.335 & 10 & 14 &  13 & 185 & 106 & 219 & B & B & \\
C002 & 184.62722 & 47.33035 &  6.980 & 25.259 & 25.181 & 25.828 & 17 & 25 &  26 & 229 & 181 & 380 & A & A & \\
C003 & 184.86600 & 47.24885 &  7.192 & 25.607 & 25.501 & 26.079 & 22 & 40 &  26 & 214 & 175 & 335 & A & A & \\
C004 & 184.71477 & 47.38594 &  7.212 & 25.330 & 25.164 & 25.755 & 23 & 28 &  31 & 332 & 245 & 486 & A & A & \\
C005 & 184.87254 & 47.23393 &  7.441 & 25.115 & 25.009 & 25.752 & 13 & 23 &  19 & 237 & 202 & 389 & A & A & \\
C006 & 184.86076 & 47.16932 &  8.022 & 25.326 & 25.176 &    \nd & 17 & 27 & \nd & 295 & 234 & \nd & A & A & \\
C007 & 184.61502 & 47.35720 &  8.075 & 25.321 & 25.307 & 25.589 & 22 & 35 &  28 & 344 & 222 & 557 & A & A & \\
C008 & 184.69374 & 47.34410 &  8.076 & 25.100 & 24.628 &    \nd & 22 & 20 & \nd & 196 & 141 & \nd & A & A & MI118600 \\
C009 & 184.82037 & 47.16744 &  8.285 & 25.114 & 24.975 & 25.415 & 14 & 22 &  20 & 190 & 132 & 385 & B & A & \\
C010 & 184.83517 & 47.21822 &  8.327 & 25.614 & 24.731 &    \nd & 26 & 21 & \nd & 334 & 217 & \nd & A & B & \\
C011 & 184.85062 & 47.19240 &  8.333 & 25.142 & 25.252 & 25.409 & 17 & 33 &  22 & 230 & 183 & 398 & A & A & MO011990\\
C012 & 184.87982 & 47.18283 &  8.520 & 25.227 & 24.358 &    \nd & 19 & 14 & \nd & 274 & 221 & \nd & A & B & \\
C013 & 184.86028 & 47.17903 &  8.588 & 25.095 & 24.745 & 25.603 & 14 & 18 &  26 & 194 & 125 & 392 & B & A & MO005306\\
C014 & 184.84859 & 47.19366 &  8.588 & 24.816 & 24.564 & 25.150 & 12 & 16 &  16 & 248 & 191 & 344 & A & A & \\
C015 & 184.84962 & 47.22719 &  8.592 & 25.202 & 24.586 &    \nd & 19 & 17 & \nd & 271 & 162 & \nd & B & B & \\
C016 & 184.64165 & 47.32391 &  8.610 & 24.851 & 24.665 & 25.263 & 12 & 17 &  16 & 259 & 198 & 274 & B & A & \\
C017 & 184.62840 & 47.38239 &  8.861 & 24.683 & 24.396 & 25.269 & 11 & 14 &  17 & 244 & 151 & 420 & B & A & \\
C018 & 184.70121 & 47.39193 &  9.102 & 25.058 & 24.504 &    \nd & 16 & 14 & \nd & 242 & 179 & \nd & A & B & \\
C019 & 184.85887 & 47.20052 &  9.135 & 24.777 & 24.278 &    \nd & 12 & 14 & \nd & 208 & 129 & \nd & B & B & MO009786\\
C020 & 184.61926 & 47.34332 &  9.205 & 25.368 & 25.239 & 25.809 & 19 & 29 &  27 & 239 & 192 & 368 & A & A & \\
C021 & 184.83388 & 47.20322 &  9.282 & 25.364 & 24.860 & 26.055 & 21 & 23 &  56 & 272 & 244 & 511 & A & A & MO025226\\
C022 & 184.79720 & 47.24684 &  9.319 & 24.980 & 24.547 & 25.778 & 14 & 15 &  23 & 166 & 117 & 323 & A & A & \\
C023 & 184.84171 & 47.23315 &  9.353 & 25.402 & 24.791 &    \nd & 18 & 19 & \nd & 171 & 163 & \nd & B & B & \\
C024 & 184.82768 & 47.24281 &  9.376 & 24.842 & 24.662 & 25.513 & 12 & 17 &  17 & 373 & 316 & 533 & A & A & \\
C025 & 184.63615 & 47.35587 &  9.546 & 25.172 & 24.755 &    \nd & 17 & 19 & \nd & 216 & 154 & \nd & A & A & \\
C026 & 184.79710 & 47.20031 &  9.633 & 24.628 & 24.355 & 24.902 & 12 & 16 &  14 & 279 & 204 & 325 & B & A & \\
C027 & 184.64081 & 47.40509 &  9.833 & 25.214 & 25.105 & 25.782 & 21 & 27 &  33 & 366 & 228 & 563 & B & A & \\
C028 & 184.80280 & 47.17722 &  9.907 & 25.259 & 24.834 &    \nd & 19 & 24 & \nd & 188 & 173 & \nd & A & A & \\
C029 & 184.79607 & 47.20068 &  9.929 & 24.388 & 24.379 &    \nd & 11 & 16 & \nd & 148 & 135 & \nd & A & A & \\
C030 & 184.82851 & 47.23561 & 10.210 & 25.518 & 24.998 &    \nd & 21 & 22 & \nd & 321 & 306 & \nd & B & A & \\
C031 & 184.70499 & 47.33163 & 10.724 & 24.578 & 24.537 & 24.823 & 13 & 19 &  16 & 278 & 229 & 299 & B & A & MI091209\\
C032 & 184.83676 & 47.17423 & 10.950 & 24.952 & 24.753 &    \nd & 12 & 17 & \nd & 264 & 204 & \nd & A & A & MO014709\\
C033 & 184.74165 & 47.36098 & 11.033 & 24.704 & 23.999 &    \nd & 12 & 09 & \nd & 255 & 170 & \nd & A & B & \\
C034 & 184.65521 & 47.32207 & 11.082 & 24.841 & 24.474 & 25.469 & 12 & 12 &  19 & 191 & 136 & 276 & A & A & \\
C035 & 184.62136 & 47.34912 & 11.503 & 24.996 & 25.064 & 25.484 & 13 & 25 &  20 & 335 & 308 & 516 & A & A & \\
C036 & 184.69243 & 47.35635 & 12.006 & 25.140 & 24.677 & 25.946 & 24 & 24 &  55 & 285 & 170 & 481 & B & A & MI126353\\
C037 & 184.65205 & 47.37317 & 12.589 & 25.475 & 24.988 &    \nd & 27 & 27 & \nd & 386 & 263 & \nd & A & A & \\
C038 & 184.84975 & 47.16279 & 13.020 & 24.853 & 24.586 & 25.156 & 12 & 15 &  16 & 359 & 324 & 571 & A & A & \\
C039 & 184.74147 & 47.36389 & 13.091 & 25.106 & 24.682 & 25.682 & 20 & 19 &  30 & 307 & 240 & 437 & A & A & \\
C040 & 184.65503 & 47.39910 & 13.490 & 24.230 & 24.022 &    \nd & 08 & 09 & \nd & 265 & 180 & \nd & A & A & \\
C041 & 184.78815 & 47.23692 & 13.715 & 25.114 & 24.835 &    \nd & 17 & 23 & \nd & 290 & 242 & \nd & A & A & \\
C042 & 184.85266 & 47.17209 & 13.896 & 24.592 & 24.384 & 24.800 & 10 & 15 &  12 & 293 & 263 & 483 & A & A & \\
C043 & 184.61859 & 47.35511 & 14.008 & 24.476 & 24.484 & 24.929 & 10 & 15 &  12 & 348 & 299 & 430 & A & A & \\
C044 & 184.69910 & 47.35472 & 15.112 & 24.230 & 24.191 & 24.579 & 09 & 13 &  12 & 416 & 338 & 453 & B & A & MI117710\\
C045 & 184.65344 & 47.32998 & 15.585 & 24.496 & 24.462 & 25.248 & 09 & 14 &  17 & 351 & 263 & 526 & A & A & MI144134\\
C046 & 184.83560 & 47.17372 & 15.801 & 24.409 & 24.196 & 24.783 & 07 & 11 &  12 & 296 & 259 & 499 & A & A & MO015276\\
C047 & 184.66770 & 47.33579 & 15.931 & 24.620 & 24.234 & 25.514 & 13 & 14 &  30 & 472 & 282 & 920 & B & A & MI138294\\
C048 & 184.70912 & 47.32398 & 16.778 & 24.471 & 24.066 & 25.508 & 15 & 13 &  50 & 387 & 287 & 776 & B & A & MI075254\\
C049 & 184.83369 & 47.24899 & 16.790 & 24.148 & 23.830 & 24.850 & 06 & 08 &  09 & 301 & 233 & 441 & A & A & \\
C050 & 184.83134 & 47.16879 & 16.975 & 24.112 & 23.935 & 24.342 & 05 & 08 &  08 & 345 & 228 & 473 & A & A & F56\\
C051 & 184.70419 & 47.37639 & 17.560 & 24.469 & 24.522 & 25.079 & 11 & 17 &  15 & 423 & 302 & 453 & B & A & \\
C052 & 184.69827 & 47.35901 & 17.567 & 24.404 & 24.292 & 25.187 & 10 & 14 &  20 & 389 & 271 & 545 & A & A & MI121312\\
C053 & 184.85165 & 47.22660 & 18.317 & 24.205 & 23.904 & 24.769 & 07 & 09 &  12 & 453 & 325 & 711 & A & A & \\
C054 & 184.71974 & 47.36516 & 18.533 & 24.473 & 24.497 & 25.147 & 12 & 18 &  19 & 443 & 385 & 635 & A & A & \\
C055 & 184.85487 & 47.19178 & 19.219 & 24.065 & 23.655 & 24.701 & 06 & 07 &  11 & 333 & 214 & 544 & A & A & \\
C056 & 184.80025 & 47.20615 & 21.265 & 24.573 & 24.280 & 25.473 & 12 & 15 &  34 & 450 & 401 & 852 & A & A & \\
C057 & 184.69022 & 47.33243 & 21.919 & 24.377 & 24.244 &    \nd & 12 & 13 & \nd & 398 & 332 & \nd & A & A & MI116159\\
C058 & 184.78761 & 47.17361 & 22.093 & 24.002 & 23.780 & 24.368 & 05 & 07 &  09 & 330 & 293 & 494 & A & A & F48\\
C059 & 184.69841 & 47.33317 & 22.651 & 23.475 & 23.096 & 24.322 & 06 & 06 &  11 & 356 & 207 & 691 & B & B & MI104131,F40\\
C060 & 184.68451 & 47.39373 & 23.803 & 23.945 & 23.819 & 24.750 & 05 & 07 &  11 & 397 & 315 & 678 & A & A & \\
C061 & 184.70107 & 47.33777 & 24.417 & 23.441 & 23.742 & 24.270 & 10 & 16 &  18 & 330 & 300 & 552 & A & B & MI103070\\
C062 & 184.85942 & 47.24532 & 24.599 & 24.145 & 23.851 & 24.812 & 09 & 10 &  23 & 434 & 289 & 868 & B & A & F35\\
C063 & 184.68946 & 47.32550 & 25.585 & 24.512 & 23.889 &    \nd & 11 & 08 & \nd & 403 & 274 & \nd & A & A & \\
C064 & 184.69884 & 47.35583 & 25.790 & 23.654 & 23.386 & 23.885 & 06 & 07 &  06 & 323 & 298 & 639 & A & A & MI118782,F14\\
C065 & 184.78604 & 47.23080 & 27.196 & 24.365 & 23.834 &    \nd & 17 & 18 & \nd & 330 & 316 & \nd & B & A & \\
C066 & 184.79041 & 47.18555 & 29.487 & 23.507 & 23.236 & 23.832 & 05 & 07 &  07 & 432 & 316 & 700 & A & A & \\
C067 & 184.70006 & 47.40289 & 31.053 & 23.377 & 23.266 & 24.045 & 03 & 05 &  06 & 402 & 300 & 655 & A & A & F22\\
C068 & 184.80029 & 47.20732 & 31.192 & 23.970 & 23.621 & 24.640 & 06 & 07 &  11 & 436 & 343 & 653 & A & A & F07\\
C069 & 184.62033 & 47.33033 & 32.250 & 23.747 & 23.606 & 24.534 & 05 & 06 &  09 & 455 & 359 & 693 & A & A & F44\\
C070 & 184.85760 & 47.22968 & 32.302 & 23.945 & 23.511 & 24.875 & 05 & 06 &  08 & 378 & 257 & 552 & A & A & F04\\
C071 & 184.85500 & 47.16898 & 32.784 & 23.744 & 23.375 & 24.190 & 05 & 06 &  07 & 438 & 315 & 693 & A & A & MO005713\\
C072 & 184.85625 & 47.16104 & 33.148 & 23.811 & 23.484 & 24.327 & 03 & 05 &  07 & 307 & 259 & 428 & A & A & F51\\
C073 & 184.73065 & 47.31969 & 33.662 & 22.848 & 22.698 & 23.478 & 05 & 06 &  08 & 305 & 268 & 430 & A & A & MI008723\\
C074 & 184.87186 & 47.22579 & 33.943 & 23.503 & 23.181 & 24.270 & 03 & 04 &  06 & 455 & 353 & 727 & A & A & F17\\
C075 & 184.61339 & 47.35845 & 34.991 & 22.928 & 22.853 & 23.386 & 02 & 03 &  03 & 369 & 302 & 432 & B & A & \\
C076 & 184.71275 & 47.35470 & 36.981 & 23.630 & 23.879 & 24.460 & 06 & 12 &  13 & 457 & 382 & 631 & A & B & MI095995,F31\\
C077 & 184.72058 & 47.39204 & 38.684 & 22.947 & 22.920 & 23.594 & 02 & 03 &  04 & 426 & 319 & 674 & A & A & F09\\
C078 & 184.79111 & 47.18379 & 38.760 & 23.211 & 22.873 & 23.725 & 02 & 02 &  06 & 469 & 329 & 762 & A & A & 
\enddata
\tablecomments{\it Table continues in next page.}
\end{deluxetable}

\clearpage

\addtocounter{table}{-1}
\begin{deluxetable}{lllrrrrrrrrrrccl}
\tablecaption{\ngal\ Cepheids -- {\it continued}}
\tabletypesize{\tiny}
\tablewidth{\textwidth}
\tablehead{
\colhead{ID} & \colhead{RA} & \colhead {Dec} & \colhead{P}   & \multicolumn{6}{c}{Mean magnitudes} & \multicolumn{3}{c}{Lightcurve ampl.} & \multicolumn{2}{c}{Qual.} & \colhead{Cross}\\
\colhead{}   & \multicolumn{2}{c}{(J2000)}   & \colhead{}    & \colhead{$r$} & \colhead{$i$} & \colhead{$g$}  & \colhead{$\sigma_r$} & \colhead{$\sigma_i$} & \colhead{$\sigma_g$} & \colhead{$r$} & \colhead{$i$} & \colhead{$g$} & \multicolumn{2}{c}{flag} & \colhead{ID}   \\
\colhead{}   & \multicolumn{2}{c}{[deg]}     & \colhead{[d]} & \multicolumn{3}{c}{[mag]} & \multicolumn{3}{c}{[mmag]} & \multicolumn{3}{c}{[mmag]} & \colhead{A} & \colhead{R} & }
\startdata
C079 & 184.73595 & 47.39787 & 39.108 & 23.189 & 23.085 & 23.998 & 02 & 03 &  05 & 392 & 303 & 680 & A & A & F23\\
C080 & 184.73083 & 47.33796 & 40.951 & 22.865 & 22.802 & 23.523 & 05 & 06 &  07 & 374 & 309 & 450 & B & A & MI032759\\
C081 & 184.70659 & 47.32061 & 44.551 & 23.005 & 22.950 & 23.522 & 03 & 04 &  06 & 352 & 304 & 457 & A & A & MI077610\\
C082 & 184.84634 & 47.24652 & 45.562 & 23.027 & 22.690 & 23.834 & 02 & 03 &  03 & 296 & 212 & 461 & A & A & F18\\
C083 & 184.71928 & 47.34868 & 47.104 & 23.363 & 23.068 &    \nd & 06 & 06 & \nd & 300 & 255 & \nd & A & A & F64\\
C084 & 184.72884 & 47.37802 & 51.629 & 23.662 & 23.380 & 24.485 & 04 & 06 &  07 & 214 & 137 & 331 & A & A & F24\\
C085 & 184.71481 & 47.30915 & 51.896 & 23.282 & 23.081 & 23.842 & 07 & 07 &  08 & 258 & 183 & 402 & A & A & \\
C086 & 184.70316 & 47.31416 & 57.338 & 22.661 & 22.096 &    \nd & 03 & 02 & \nd & 270 & 218 & \nd & A & A & \\
C087 & 184.85777 & 47.16559 & 58.984 & 22.372 & 21.732 & 23.656 & 02 & 02 &  04 & 144 & 109 & 149 & B & B & \\
C088 & 184.72849 & 47.31558 & 78.078 & 22.560 & 22.048 &    \nd & 06 & 07 & \nd & 179 & 116 & \nd & A & A & \\
C089 & 184.84549 & 47.21271 & 83.258 & 22.616 & 22.257 & 23.478 & 02 & 02 &  03 & 155 & 144 & 242 & B & A & \\
C090 & 184.84308 & 47.17088 &100.297 & 22.208 & 21.485 & 23.661 & 01 & 01 &  04 & 145 & 104 & 203 & A & B & \\
C091 & 184.69186 & 47.38691 &105.750 & 21.667 & 21.429 & 22.146 & 01 & 01 &  01 & 236 & 173 & 321 & A & A & \\
C092 & 184.85767 & 47.16682 &109.365 & 21.753 & 21.424 & 22.347 & 01 & 01 &  02 & 196 & 157 & 291 & A & A & \\
C093 & 184.81371 & 47.19359 &115.654 & 22.497 & 21.902 &    \nd & 02 & 02 & \nd & 172 & 129 & \nd & A & A & \\
C094 & 184.73400 & 47.32057 &127.408 & 22.732 & 21.881 &    \nd & 05 & 02 & \nd & 145 & 113 & \nd & A & B &
\enddata
\tablecomments{The uncertainties in mean magnitude reflect only the statistical component; please refer to Table~\ref{tab:photcal} for systematic uncertainties. Quality flags: A, amplitude ratios; R, P-L residuals. Cross-IDs: F=\citet{fausnaugh14}; M=\citet{macri06}.}
\end{deluxetable}

\begin{figure}[h]
\begin{center}
\includegraphics[width=0.49\textwidth]{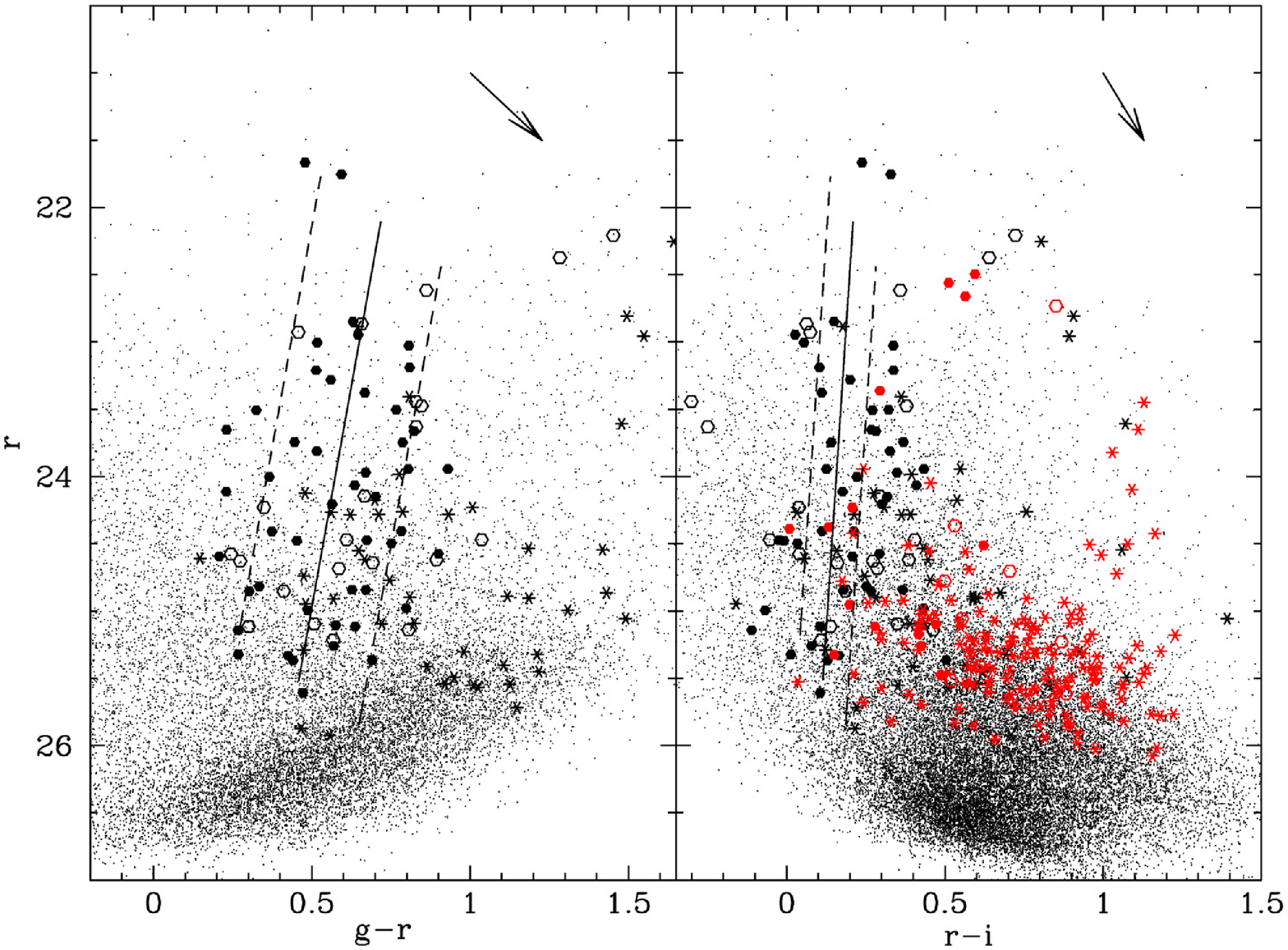}
\end{center}
\caption{Color-magnitude diagrams of stars in \ngal, using $g\mm r$ (left) and $r\mm i$ (right). Symbols have the same meaning as Fig.~8, except that red starred symbols are used to plot objects listed in Table~\ref{tab:vars}. The center of the zero-extinction instability strip is marked with a solid line, while the dotted lines represent its $2\sigma$ width. Extinction vectors for $A_r\ig 0.2$~mag are plotted.\label{fig:cmd}}
\end{figure}

\noindent{ing the magnitude difference between the \citet{kodric13} ``PLC'' relations for classical (``FM'') and Population II (``T2'') Cepheids in $r$ and $i$ for periods ranging from $30\mm100$d, which exhibited a dispersion of only 0.02~mag. Color-magnitude diagrams of the Cepheids and other variables are plotted in Figure~\ref{fig:cmd}. The semi-empirical P-L relations of \S\ref{SC:SLOANPL} were used to illustrate the approximate location and intrinsic width of the zero-extinction instability strip. There is some evidence for differential extinction among Cepheids with $P\!>\!80$d, which is commonly seen in other galaxies since these are the youngest Cepheids and therefore are closest to their natal regions. The variables listed in Table~\ref{tab:vars} are mostly located in the AGB/RGB region of the diagram, as expected given their likely nature (Population II pulsator or highly-reddened classical Cepheid).}

\ \par

\ \par

\ \par

\vspace{2.71in}

\section{Prospects for LSST\label{sc:lsst}}

Our results have demonstrated the feasibility of discovering Cepheids and other long-period variables with 8-m class telescopes out to significantly larger distances than before \citep[$D\sim4.5$~Mpc for M83,][]{thim03}. Furthermore, the work carried out by \citet{gerke11} and \citet{fausnaugh14} have highlighted the efficacy of difference imaging techniques for these surveys, as originally demonstrated by \citet{bonanos03}.

\vspace{3pt}

The Large Synoptic Survey Telescope (LSST), slated to start operations by the end of the decade, will deliver images of most of the southern sky with angular resolution and depth ($5\sigma$ limiting magnitude) comparable to the data collected as part of our survey (LSST average values: $0\farcs73$, $g\sim 24.9$, $r\sim 24.6$, $i\sim 24.0$; our survey: $<0\farcs7$, $g \sim 26.5$, $r \sim 26.4$, $i \sim 25.8$), but with a vastly superior temporal sampling (LSST: $\sim 32$ epochs in $g$ and $\sim 73$ in $r$ \& $i$; our survey: $\sim 16$ per band). Based on the calculations described below, we expect that LSST will enable efficient searches for Cepheids and long-period variables in a considerable number of galaxies out to at least $D\sim 10$~Mpc. At this distance, the typical LSST single-image depth in $r$ will be comparable to the mean magnitude of a $P\sim25$d classical Cepheid or a $P\sim100$d Pop II variable.

\vspace{3pt}

We used the Extragalactic Distance Database \citep[EDD,][]{tully09} and the Cosmicflows-2 catalog of distances \citep{tully13} to identify spiral or dwarf galaxies that would be suitable for Cepheid searches with LSST based on the following criteria: (i) $D<10$~Mpc; (ii) $-63^\circ < \delta < 0^\circ$ and $|b|\!\geq\!10^\circ$ (the approximate boundaries of the ``wide-fast-deep'' survey mode); (iii) $i\leq78^\circ$ for spirals (i.e., no more inclined than M31). There are \nlsst\ galaxies that meet this criteria, which are listed in Table~\ref{tab:lsstgal}. We include all dwarf galaxies regardless of their recent star formation history because Population II pulsators should be detectable (with a period limit $\sim 4\times$ larger than Population I Cepheids for a given apparent magnitude limit). We also included NGC$\,$5128 despite its ``early type'' classification because it has been shown to host Population I Cepheids \citep{ferrarese07} as well as a significant population of LPVs \citep{rejkuba04}.

\clearpage

\begin{deluxetable}{lllrrrrrrrrrrrrrccr}
\tablecaption{\ngal\ Variables \label{tab:vars}}
\tablewidth{\textwidth}
\tabletypesize{\tiny}
\tablehead{
\colhead{ID} & \colhead{RA} & \colhead {Dec} & \colhead{P}   & \multicolumn{6}{c}{Mean magnitudes} & \multicolumn{3}{c}{Lightcurve ampl.} & \multicolumn{2}{c}{Qual.} & \colhead{Cross}\\
\colhead{}   & \multicolumn{2}{c}{(J2000)}   & \colhead{}    & \colhead{$r$} & \colhead{$i$} & \colhead{$g$}  & \colhead{$\sigma_r$} & \colhead{$\sigma_i$} & \colhead{$\sigma_g$} & \colhead{$r$} & \colhead{$i$} & \colhead{$g$} & \multicolumn{2}{c}{flag} & \colhead{ID}   \\
\colhead{}   & \multicolumn{2}{c}{[deg]}     & \colhead{[d]} & \multicolumn{3}{c}{[mag]} & \multicolumn{3}{c}{[mmag]} & \multicolumn{3}{c}{[mmag]} & \colhead{A} & \colhead{R} & }
\startdata
V001 & 184.61588 & 47.35870 &   7.208 & 25.526 & 25.490 &    \nd & 27 & 45 &\nd & 275 & 277 &  \nd & C & A & \\
V002 & 184.73772 & 47.34910 &   7.482 & 24.928 & 24.617 &    \nd & 18 & 20 &\nd & 211 & 107 &  \nd & C & A & \\
V003 & 184.63512 & 47.36143 &   7.643 & 24.945 & 25.105 & 25.426 & 15 & 27 & 19 & 277 & 197 &  268 & C & A & \\
V004 & 184.71381 & 47.35652 &   7.862 & 24.740 & 24.493 & 25.215 & 17 & 20 & 28 & 315 & 161 &  478 & C & A & MI095711\\
V005 & 184.64999 & 47.32576 &   8.956 & 24.552 & 24.394 & 25.200 & 09 & 12 & 14 & 259 & 138 &  315 & C & A & MI144791\\
V006 & 184.83194 & 47.24321 &  10.278 & 25.060 & 24.311 &    \nd & 13 & 12 &\nd & 241 & 135 &  \nd & C & B & \\
V007 & 184.79041 & 47.20185 &  11.212 & 24.609 & 24.553 & 24.756 & 12 & 18 & 13 & 287 & 209 &  283 & C & A & \\
V008 & 184.70895 & 47.33271 &  11.472 & 24.912 & 24.323 & 25.481 & 18 & 15 & 32 & 403 & 208 &  645 & C & B & MI084547\\
V009 & 184.78964 & 47.23041 &  11.827 & 25.212 & 24.913 &    \nd & 18 & 25 &\nd & 279 & 295 &  \nd & C & A & \\
V010 & 184.82361 & 47.16617 &  12.068 & 26.026 & 25.050 &    \nd & 37 & 22 &\nd & 476 & 293 &  \nd & B & C & \\
V011 & 184.68954 & 47.33534 &  12.493 & 25.094 & 24.704 & 25.815 & 19 & 20 & 37 & 339 & 180 &  396 & C & A & \\
V012 & 184.64534 & 47.33925 &  12.517 & 25.719 & 25.497 & 26.868 & 30 & 41 & 83 & 391 & 365 &  800 & B & C & \\
V013 & 184.87843 & 47.16182 &  12.716 & 25.721 & 24.951 &    \nd & 23 & 20 &\nd & 270 & 266 &  \nd & B & C & \\
V014 & 184.82854 & 47.23107 &  13.132 & 25.910 & 24.989 &    \nd & 45 & 28 &\nd & 437 & 331 &  \nd & A & C & \\
V015 & 184.85751 & 47.20291 &  14.208 & 25.670 & 24.833 &    \nd & 26 & 22 &\nd & 361 & 297 &  \nd & A & C & \\
V016 & 184.69113 & 47.34583 &  14.379 & 23.985 & 23.590 & 24.761 & 07 & 08 & 16 & 355 & 183 &  673 & C & B & MI122858\\
V017 & 184.73586 & 47.35723 &  15.055 & 24.122 & 23.846 & 24.601 & 08 & 11 & 12 & 360 & 191 &  481 & C & A & MI043585\\
V018 & 184.63998 & 47.39553 &  15.526 & 25.553 & 25.201 & 26.565 & 30 & 30 & 83 & 486 & 285 &  407 & C & C & \\
V019 & 184.75009 & 47.36823 &  16.992 & 25.465 & 24.954 &    \nd & 30 & 27 &\nd & 656 & 373 &  \nd & B & C & \\
V020 & 184.74911 & 47.37642 &  18.072 & 25.621 & 25.234 &    \nd & 27 & 27 &\nd & 380 & 214 &  \nd & C & C & \\
V021 & 184.64842 & 47.37423 &  19.681 & 25.290 & 24.873 &    \nd & 23 & 26 &\nd & 424 & 291 &  \nd & A & C & \\
V022 & 184.65116 & 47.34100 &  19.827 & 25.645 & 24.705 &    \nd & 32 & 18 &\nd & 675 & 485 &  \nd & A & C & \\
V023 & 184.69412 & 47.39306 &  19.849 & 25.705 & 24.903 &    \nd & 35 & 22 &\nd & 555 & 411 &  \nd & A & C & \\
V024 & 184.69690 & 47.31952 &  19.897 & 25.312 & 24.529 &    \nd & 25 & 18 &\nd & 304 & 271 &  \nd & A & C & \\
V025 & 184.72200 & 47.37108 &  19.923 & 24.996 & 24.569 & 26.305 & 19 & 19 & 62 & 468 & 314 & 1059 & C & C & \\
V026 & 184.69537 & 47.34772 &  21.008 & 25.261 & 24.561 &    \nd & 30 & 22 &\nd & 421 & 384 &  \nd & A & C & \\
V027 & 184.73750 & 47.36486 &  21.229 & 25.334 & 24.619 &    \nd & 27 & 22 &\nd & 387 & 294 &  \nd & A & C & \\
V028 & 184.68947 & 47.33310 &  21.288 & 25.290 & 25.167 & 25.765 & 29 & 37 & 38 & 426 & 288 &  688 & A & C & MI117637\\
V029 & 184.65191 & 47.33193 &  21.400 & 25.774 & 24.944 &    \nd & 37 & 23 &\nd & 533 & 321 &  \nd & B & C & \\
V030 & 184.72806 & 47.39740 &  21.820 & 25.725 & 25.179 &    \nd & 33 & 27 &\nd & 466 & 331 &  \nd & A & C & \\
V031 & 184.69263 & 47.37821 &  22.316 & 25.311 & 24.552 &    \nd & 23 & 17 &\nd & 365 & 300 &  \nd & A & C & \\
V032 & 184.67262 & 47.34022 &  22.545 & 25.529 & 24.416 &    \nd & 28 & 13 &\nd & 618 & 383 &  \nd & B & C & \\
V033 & 184.87793 & 47.23467 &  22.808 & 25.765 & 24.613 &    \nd & 29 & 16 &\nd & 536 & 395 &  \nd & A & C & \\
V034 & 184.81133 & 47.17447 &  23.229 & 25.451 & 24.618 &    \nd & 22 & 17 &\nd & 469 & 323 &  \nd & A & C & \\
V035 & 184.67284 & 47.38055 &  23.541 & 25.026 & 24.556 &    \nd & 17 & 17 &\nd & 470 & 319 &  \nd & A & C & \\
V036 & 184.68389 & 47.33776 &  24.043 & 25.172 & 24.207 &    \nd & 23 & 13 &\nd & 658 & 621 &  \nd & B & C & \\
V037 & 184.74284 & 47.38165 &  24.165 & 25.851 & 25.260 &    \nd & 37 & 28 &\nd & 433 & 311 &  \nd & A & C & \\
V038 & 184.85094 & 47.23927 &  24.460 & 26.078 & 24.922 &    \nd & 43 & 28 &\nd & 399 & 373 &  \nd & B & C & \\
V039 & 184.63644 & 47.37076 &  24.610 & 25.239 & 24.866 &    \nd & 20 & 22 &\nd & 380 & 352 &  \nd & A & C & \\
V040 & 184.67960 & 47.36776 &  25.304 & 25.140 & 24.349 &    \nd & 21 & 14 &\nd & 576 & 305 &  \nd & C & C & \\
V041 & 184.68857 & 47.35267 &  25.439 & 25.419 & 24.484 &    \nd & 38 & 21 &\nd & 492 & 478 &  \nd & B & C & \\
V042 & 184.80403 & 47.25509 &  25.604 & 25.340 & 24.359 &    \nd & 19 & 13 &\nd & 482 & 290 &  \nd & B & C & \\
V043 & 184.73220 & 47.38623 &  25.620 & 25.690 & 24.969 &    \nd & 32 & 23 &\nd & 482 & 479 &  \nd & B & C & \\
V044 & 184.86981 & 47.16333 &  25.863 & 25.621 & 24.769 &    \nd & 26 & 18 &\nd & 597 & 442 &  \nd & A & C & \\
V045 & 184.65614 & 47.37715 &  25.990 & 25.321 & 24.476 &    \nd & 22 & 17 &\nd & 438 & 353 &  \nd & A & C & \\
V046 & 184.66753 & 47.35943 &  26.103 & 25.552 & 24.961 &    \nd & 31 & 26 &\nd & 493 & 399 &  \nd & A & C & \\
V047 & 184.78963 & 47.22095 &  26.125 & 25.225 & 24.276 &    \nd & 25 & 14 &\nd & 593 & 376 &  \nd & B & C & \\
V048 & 184.79379 & 47.24938 &  26.218 & 25.336 & 24.755 &    \nd & 27 & 22 &\nd & 813 & 625 &  \nd & A & C & \\
V049 & 184.68507 & 47.34696 &  26.414 & 25.419 & 24.367 &    \nd & 33 & 17 &\nd & 534 & 339 &  \nd & B & C & \\
V050 & 184.66936 & 47.40280 &  26.460 & 25.548 & 24.911 &    \nd & 30 & 22 &\nd & 504 & 365 &  \nd & A & C & \\
V051 & 184.62876 & 47.33155 &  26.544 & 25.679 & 25.437 &    \nd & 33 & 43 &\nd & 394 & 388 &  \nd & B & C & \\
V052 & 184.71328 & 47.39986 &  26.962 & 25.830 & 24.936 &    \nd & 34 & 20 &\nd & 359 & 237 &  \nd & A & C & \\
V053 & 184.82440 & 47.22630 &  27.004 & 25.568 & 25.058 & 26.594 & 43 & 41 & 83 & 469 & 464 &  503 & B & C & \\
V054 & 184.89491 & 47.23735 &  27.030 & 25.874 & 25.658 & 26.340 & 37 & 65 & 42 & 414 & 315 &  589 & A & C & \\
V055 & 184.68645 & 47.36620 &  27.285 & 25.207 & 24.576 &    \nd & 25 & 20 &\nd & 674 & 400 &  \nd & B & C & \\
V056 & 184.89307 & 47.20862 &  27.488 & 25.926 & 25.217 & 26.483 & 41 & 33 & 83 & 459 & 284 &  486 & B & C & \\
V057 & 184.88348 & 47.19092 &  27.874 & 25.451 & 24.832 & 26.669 & 24 & 21 & 83 & 554 & 451 &  500 & C & C & \\
V058 & 184.73781 & 47.37998 &  27.908 & 25.543 & 24.941 & 26.464 & 26 & 20 & 42 & 424 & 285 &  391 & C & C & \\
V059 & 184.64410 & 47.35738 &  28.308 & 25.694 & 25.268 &    \nd & 36 & 33 &\nd & 349 & 315 &  \nd & A & C & \\
V060 & 184.72662 & 47.36694 &  28.310 & 25.507 & 24.689 &    \nd & 37 & 20 &\nd & 693 & 357 &  \nd & C & C & \\
V061 & 184.87186 & 47.22157 &  28.713 & 25.573 & 25.273 &    \nd & 31 & 33 &\nd & 493 & 292 &  \nd & B & C & \\
V062 & 184.63959 & 47.35273 &  28.719 & 25.652 & 24.594 &    \nd & 34 & 17 &\nd & 548 & 355 &  \nd & A & C & \\
V063 & 184.81345 & 47.20218 &  28.756 & 25.415 & 24.453 &    \nd & 23 & 15 &\nd & 304 & 289 &  \nd & B & C & \\
V064 & 184.68690 & 47.39297 &  28.928 & 24.898 & 24.615 & 25.708 & 15 & 19 & 29 & 348 & 243 &  619 & A & C & \\
V065 & 184.67148 & 47.35087 &  29.489 & 25.110 & 24.697 &    \nd & 24 & 27 &\nd & 387 & 356 &  \nd & A & C & \\
V066 & 184.72989 & 47.36077 &  29.676 & 24.935 & 24.057 &    \nd & 23 & 23 &\nd & 416 & 221 &  \nd & C & C & \\
V067 & 184.83688 & 47.24517 &  29.814 & 25.845 & 25.048 &    \nd & 32 & 24 &\nd & 396 & 267 &  \nd & A & C & \\
V068 & 184.80092 & 47.23567 &  30.356 & 25.413 & 25.015 & 26.277 & 27 & 34 & 64 & 380 & 228 &  624 & B & C & \\
V069 & 184.73583 & 47.38627 &  30.446 & 25.295 & 24.654 &    \nd & 22 & 15 &\nd & 394 & 366 &  \nd & B & C & \\
V070 & 184.62962 & 47.37225 &  30.642 & 25.572 & 24.513 &    \nd & 34 & 18 &\nd & 527 & 429 &  \nd & A & C & \\
V071 & 184.72589 & 47.37442 &  31.153 & 25.350 & 24.268 &    \nd & 27 & 13 &\nd & 392 & 333 &  \nd & A & C & \\
V072 & 184.85495 & 47.16732 &  31.635 & 25.469 & 25.257 &    \nd & 28 & 41 &\nd & 288 & 306 &  \nd & C & C & MO005461\\
V073 & 184.83421 & 47.22531 &  31.918 & 24.994 & 24.563 &    \nd & 14 & 16 &\nd & 335 & 283 &  \nd & A & C & \\
V074 & 184.64639 & 47.32715 &  32.022 & 25.416 & 24.841 &    \nd & 25 & 19 &\nd & 500 & 411 &  \nd & A & C & \\
V075 & 184.68745 & 47.35280 &  32.423 & 25.005 & 24.394 &    \nd & 18 & 17 &\nd & 421 & 395 &  \nd & B & C & \\
V076 & 184.63246 & 47.38546 &  32.580 & 25.478 & 24.980 &    \nd & 28 & 27 &\nd & 449 & 276 &  \nd & B & C & \\
V077 & 184.81525 & 47.23255 &  32.590 & 25.531 & 24.957 &    \nd & 28 & 31 &\nd & 526 & 483 &  \nd & A & C & \\
V078 & 184.83374 & 47.17822 &  32.647 & 25.457 & 24.487 &    \nd & 24 & 17 &\nd & 564 & 292 &  \nd & C & C &
\enddata
\tablecomments{\it Table continues in next page.}
\end{deluxetable}

\clearpage

\addtocounter{table}{-1}
\begin{deluxetable}{lllrrrrrrrrrrrrrccr}
\tablecaption{\ngal\ Variables -- {\it continued}}
\tablewidth{\textwidth}
\tabletypesize{\tiny}
\tablehead{
\colhead{ID} & \colhead{RA} & \colhead {Dec} & \colhead{P}   & \multicolumn{6}{c}{Mean magnitudes} & \multicolumn{3}{c}{Lightcurve ampl.} & \multicolumn{2}{c}{Qual.} & \colhead{Cross}\\
\colhead{}   & \multicolumn{2}{c}{(J2000)}   & \colhead{}    & \colhead{$r$} & \colhead{$i$} & \colhead{$g$}  & \colhead{$\sigma_r$} & \colhead{$\sigma_i$} & \colhead{$\sigma_g$} & \colhead{$r$} & \colhead{$i$} & \colhead{$g$} & \multicolumn{2}{c}{flag} & \colhead{ID}   \\
\colhead{}   & \multicolumn{2}{c}{[deg]}     & \colhead{[d]} & \multicolumn{3}{c}{[mag]} & \multicolumn{3}{c}{[mmag]} & \multicolumn{3}{c}{[mmag]} & \colhead{A} & \colhead{R} & }
\startdata
V079 & 184.62106 & 47.37893 &  32.837 & 25.656 & 24.960 &    \nd & 31 & 24 &\nd & 380 & 266 &  \nd & A & C & \\
V080 & 184.89517 & 47.18497 &  33.105 & 25.605 & 24.966 &    \nd & 27 & 24 &\nd & 575 & 384 &  \nd & A & C & \\
V081 & 184.65927 & 47.32368 &  33.983 & 25.621 & 24.912 &    \nd & 29 & 21 &\nd & 507 & 255 &  \nd & C & C & \\
V082 & 184.72793 & 47.38604 &  34.048 & 25.542 & 24.977 &    \nd & 24 & 21 &\nd & 336 & 304 &  \nd & A & C & \\
V083 & 184.80679 & 47.20128 &  34.145 & 25.624 & 24.862 &    \nd & 27 & 22 &\nd & 293 & 310 &  \nd & C & C & \\
V084 & 184.85439 & 47.23996 &  34.444 & 25.706 & 24.869 &    \nd & 27 & 20 &\nd & 490 & 277 &  \nd & C & C & \\
V085 & 184.68365 & 47.37157 &  35.174 & 25.502 & 24.527 &    \nd & 31 & 17 &\nd & 405 & 302 &  \nd & A & C & \\
V086 & 184.63749 & 47.40559 &  35.394 & 25.719 & 24.726 &    \nd & 37 & 22 &\nd & 562 & 560 &  \nd & B & C & \\
V087 & 184.64494 & 47.37695 &  35.451 & 25.284 & 24.662 &    \nd & 24 & 20 &\nd & 640 & 424 &  \nd & A & C & \\
V088 & 184.66160 & 47.39078 &  35.709 & 25.770 & 24.746 &    \nd & 37 & 17 &\nd & 656 & 343 &  \nd & C & C & \\
V089 & 184.64494 & 47.38647 &  35.849 & 25.049 & 24.232 &    \nd & 17 & 13 &\nd & 406 & 284 &  \nd & A & C & \\
V090 & 184.70110 & 47.35960 &  36.192 & 25.074 & 24.417 &    \nd & 20 & 17 &\nd & 371 & 300 &  \nd & A & C & \\
V091 & 184.68969 & 47.39113 &  36.433 & 25.437 & 24.605 &    \nd & 22 & 14 &\nd & 397 & 267 &  \nd & A & C & \\
V092 & 184.68663 & 47.33139 &  36.508 & 25.433 & 24.782 &    \nd & 26 & 22 &\nd & 342 & 363 &  \nd & C & C & \\
V093 & 184.61984 & 47.37281 &  36.735 & 25.439 & 24.734 &    \nd & 25 & 21 &\nd & 477 & 472 &  \nd & B & C & \\
V094 & 184.81386 & 47.16805 &  38.094 & 25.583 & 24.845 &    \nd & 23 & 21 &\nd & 546 & 444 &  \nd & A & C & \\
V095 & 184.70255 & 47.37877 &  38.311 & 25.853 & 24.964 &    \nd & 43 & 25 &\nd & 411 & 348 &  \nd & A & C & \\
V096 & 184.68149 & 47.39181 &  38.560 & 25.817 & 25.487 &    \nd & 39 & 46 &\nd & 493 & 466 &  \nd & B & C & \\
V097 & 184.86378 & 47.23680 &  38.681 & 26.018 & 24.848 &    \nd & 40 & 23 &\nd & 426 & 440 &  \nd & C & C & \\
V098 & 184.81641 & 47.21674 &  38.810 & 25.520 & 24.635 &    \nd & 33 & 24 &\nd & 446 & 434 &  \nd & B & C & \\
V099 & 184.67522 & 47.36093 &  39.948 & 25.114 & 24.564 &    \nd & 20 & 18 &\nd & 523 & 491 &  \nd & B & C & \\
V100 & 184.61584 & 47.32947 &  40.331 & 25.954 & 25.296 &    \nd & 44 & 32 &\nd & 502 & 306 &  \nd & B & C & \\
V101 & 184.66415 & 47.37246 &  40.684 & 25.328 & 24.431 &    \nd & 26 & 18 &\nd & 410 & 226 &  \nd & C & C & \\
V102 & 184.67718 & 47.33715 &  40.812 & 25.766 & 24.544 &    \nd & 39 & 18 &\nd & 585 & 474 &  \nd & A & C & \\
V103 & 184.66820 & 47.40309 &  41.746 & 25.201 & 24.434 &    \nd & 18 & 12 &\nd & 339 & 254 &  \nd & A & C & \\
V104 & 184.68365 & 47.37054 &  41.820 & 24.890 & 24.301 & 26.008 & 17 & 14 & 42 & 450 & 305 &  416 & C & C & \\
V105 & 184.73051 & 47.37608 &  42.461 & 25.288 & 24.716 &    \nd & 24 & 20 &\nd & 314 & 329 &  \nd & C & C & \\
V106 & 184.68549 & 47.40226 &  43.276 & 24.777 & 24.601 &    \nd & 12 & 17 &\nd & 361 & 211 &  \nd & B & C & \\
V107 & 184.81400 & 47.20696 &  43.672 & 25.423 & 24.435 &    \nd & 22 & 13 &\nd & 370 & 236 &  \nd & B & C & \\
V108 & 184.62269 & 47.33659 &  43.945 & 25.288 & 24.373 &    \nd & 19 & 12 &\nd & 325 & 264 &  \nd & A & C & \\
V109 & 184.80774 & 47.24590 &  44.142 & 25.066 & 24.519 &    \nd & 16 & 17 &\nd & 469 & 411 &  \nd & A & C & \\
V110 & 184.70373 & 47.31695 &  44.145 & 24.422 & 24.212 &    \nd & 15 & 17 &\nd & 355 & 317 &  \nd & A & C & \\
V111 & 184.74081 & 47.34787 &  44.511 & 24.502 & 23.424 &    \nd & 19 & 09 &\nd & 400 & 204 &  \nd & C & C & \\
V112 & 184.64447 & 47.40064 &  44.907 & 25.433 & 24.815 &    \nd & 23 & 19 &\nd & 335 & 328 &  \nd & B & C & \\
V113 & 184.82291 & 47.18021 &  44.922 & 25.835 & 25.303 &    \nd & 33 & 33 &\nd & 490 & 381 &  \nd & A & C & \\
V114 & 184.70123 & 47.38385 &  46.016 & 25.200 & 24.624 &    \nd & 18 & 17 &\nd & 241 & 223 &  \nd & A & C & \\
V115 & 184.83519 & 47.25141 &  48.035 & 25.315 & 24.378 &    \nd & 19 & 12 &\nd & 395 & 207 &  \nd & C & C & \\
V116 & 184.64465 & 47.39151 &  50.888 & 25.247 & 24.699 &    \nd & 20 & 17 &\nd & 368 & 306 &  \nd & A & C & \\
V117 & 184.72504 & 47.37707 &  50.929 & 25.309 & 24.638 &    \nd & 22 & 17 &\nd & 389 & 306 &  \nd & A & C & \\
V118 & 184.88217 & 47.22714 &  51.012 & 25.980 & 25.063 &    \nd & 33 & 23 &\nd & 517 & 279 &  \nd & C & C & \\
V119 & 184.81825 & 47.22616 &  51.099 & 25.545 & 24.481 &    \nd & 25 & 15 &\nd & 353 & 270 &  \nd & A & C & \\
V120 & 184.64674 & 47.35896 &  51.911 & 25.112 & 24.334 &    \nd & 17 & 13 &\nd & 248 & 258 &  \nd & C & C & \\
V121 & 184.71109 & 47.35273 &  52.319 & 24.096 & 23.005 &    \nd & 09 & 05 &\nd & 328 & 164 &  \nd & C & C & \\
V122 & 184.62032 & 47.34943 &  52.367 & 24.265 & 24.232 & 24.827 & 07 & 10 & 10 & 249 & 182 &  375 & A & C & \\
V123 & 184.66360 & 47.38593 &  52.458 & 24.904 & 24.302 & 26.090 & 17 & 15 & 44 & 351 & 176 &  316 & C & C & \\
V124 & 184.83891 & 47.20099 &  52.533 & 24.262 & 23.504 & 25.049 & 12 & 10 & 21 & 275 & 184 &  320 & B & C & \\
V125 & 184.68855 & 47.34049 &  52.755 & 24.561 & 23.996 &    \nd & 12 & 11 &\nd & 318 & 235 &  \nd & A & C & \\
V126 & 184.82556 & 47.19948 &  53.460 & 25.424 & 24.285 &    \nd & 22 & 13 &\nd & 301 & 215 &  \nd & A & C & \\
V127 & 184.86745 & 47.22917 &  55.180 & 25.278 & 24.449 &    \nd & 17 & 14 &\nd & 226 & 152 &  \nd & A & C & \\
V128 & 184.69830 & 47.36268 &  55.865 & 25.140 & 24.366 &    \nd & 27 & 17 &\nd & 306 & 213 &  \nd & A & C & \\
V129 & 184.67491 & 47.40297 &  56.694 & 24.691 & 24.115 &    \nd & 11 & 09 &\nd & 189 & 175 &  \nd & A & C & \\
V130 & 184.87714 & 47.21610 &  57.464 & 25.414 & 24.782 &    \nd & 20 & 19 &\nd & 297 & 271 &  \nd & A & C & \\
V131 & 184.65292 & 47.38004 &  58.623 & 25.592 & 24.807 &    \nd & 30 & 20 &\nd & 309 & 185 &  \nd & B & C & \\
V132 & 184.70101 & 47.32054 &  59.136 & 24.425 & 23.261 &    \nd & 13 & 06 &\nd & 186 & 124 &  \nd & A & C & \\
V133 & 184.80930 & 47.20495 &  59.640 & 24.919 & 24.399 &    \nd & 15 & 14 &\nd & 321 & 197 &  \nd & B & C & \\
V134 & 184.65126 & 47.39195 &  59.895 & 25.515 & 24.875 &    \nd & 25 & 19 &\nd & 319 & 265 &  \nd & A & C & \\
V135 & 184.63419 & 47.38253 &  60.243 & 25.304 & 24.379 & 26.284 & 23 & 14 & 61 & 305 & 202 &  256 & C & C & \\
V136 & 184.82658 & 47.24206 &  62.283 & 24.876 & 24.238 &    \nd & 12 & 11 &\nd & 209 & 158 &  \nd & A & C & \\
V137 & 184.62708 & 47.37817 &  62.515 & 25.139 & 24.718 &    \nd & 19 & 19 &\nd & 290 & 202 &  \nd & A & C & \\
V138 & 184.64722 & 47.33090 &  62.968 & 25.090 & 24.196 &    \nd & 15 & 10 &\nd & 226 & 166 &  \nd & A & C & \\
V139 & 184.67236 & 47.38370 &  63.299 & 25.243 & 24.483 &    \nd & 20 & 15 &\nd & 322 & 199 &  \nd & B & C & \\
V140 & 184.70583 & 47.34950 &  63.316 & 24.047 & 23.592 &    \nd & 12 & 11 &\nd & 150 & 110 &  \nd & A & C & \\
V141 & 184.64357 & 47.40007 &  64.506 & 25.323 & 24.630 & 26.535 & 20 & 15 & 62 & 271 & 230 &  263 & C & C & \\
V142 & 184.83250 & 47.23988 &  65.440 & 25.818 & 24.753 &    \nd & 30 & 18 &\nd & 355 & 203 &  \nd & B & C & \\
V143 & 184.86067 & 47.23750 &  65.990 & 25.091 & 24.653 & 25.915 & 13 & 16 & 22 & 180 & 139 &  253 & A & C & \\
V144 & 184.79779 & 47.22414 &  66.078 & 24.286 & 23.923 & 24.908 & 08 & 11 & 13 & 201 & 183 &  287 & A & C & \\
V145 & 184.80391 & 47.20254 &  67.656 & 24.283 & 24.069 & 24.995 & 08 & 12 & 14 & 188 & 171 &  217 & B & C & \\
V146 & 184.85565 & 47.22339 &  68.615 & 23.609 & 22.538 & 25.088 & 03 & 02 & 12 & 225 & 142 &  200 & C & C & \\
V147 & 184.87787 & 47.20903 &  69.116 & 24.513 & 24.129 &    \nd & 10 & 12 &\nd & 141 & 111 &  \nd & A & C & \\
V148 & 184.83575 & 47.21942 &  70.386 & 25.432 & 24.841 &    \nd & 27 & 23 &\nd & 279 & 192 &  \nd & A & C & \\
V149 & 184.66441 & 47.35379 &  71.945 & 24.941 & 24.683 &    \nd & 16 & 20 &\nd & 302 & 209 &  \nd & A & C & \\
V150 & 184.70332 & 47.32773 &  74.032 & 24.618 & 24.171 & 25.285 & 13 & 12 & 22 & 195 & 144 &  326 & A & C & MI091129\\
V151 & 184.80777 & 47.25814 &  74.776 & 24.770 & 24.316 & 25.516 & 11 & 14 & 17 & 208 & 182 &  342 & A & C & \\
V152 & 184.77460 & 47.18651 &  74.875 & 25.404 & 24.840 & 26.510 & 19 & 18 & 83 & 382 & 302 &  320 & C & C & \\
V153 & 184.61720 & 47.36103 &  77.405 & 24.536 & 24.104 & 25.721 & 10 & 11 & 23 & 105 & 108 &  161 & C & C & \\
V154 & 184.67953 & 47.32811 &  77.785 & 25.349 & 24.598 &    \nd & 22 & 17 &\nd & 200 & 201 &  \nd & C & C & \\
V155 & 184.66908 & 47.35049 &  84.354 & 25.223 & 24.508 &    \nd & 23 & 17 &\nd & 296 & 179 &  \nd & B & C & \\
V156 & 184.64259 & 47.35006 &  85.781 & 24.864 & 24.189 & 26.295 & 12 & 12 & 43 & 176 & 121 &  241 & A & C & 
\enddata
\tablecomments{\it Table continues on next page.}
\end{deluxetable}

\clearpage

\addtocounter{table}{-1}
\begin{deluxetable}{lllrrrrrrrrrrrrrccr}
\tablecaption{\ngal\ Variables -- {\it continued}}
\tablewidth{\textwidth}
\tabletypesize{\tiny}
\tablehead{
\colhead{ID} & \colhead{RA} & \colhead {Dec} & \colhead{P}   & \multicolumn{6}{c}{Mean magnitudes} & \multicolumn{3}{c}{Lightcurve ampl.} & \multicolumn{2}{c}{Qual.} & \colhead{Cross}\\
\colhead{}   & \multicolumn{2}{c}{(J2000)}   & \colhead{}    & \colhead{$r$} & \colhead{$i$} & \colhead{$g$}  & \colhead{$\sigma_r$} & \colhead{$\sigma_i$} & \colhead{$\sigma_g$} & \colhead{$r$} & \colhead{$i$} & \colhead{$g$} & \multicolumn{2}{c}{flag} & \colhead{ID}   \\
\colhead{}   & \multicolumn{2}{c}{[deg]}     & \colhead{[d]} & \multicolumn{3}{c}{[mag]} & \multicolumn{3}{c}{[mmag]} & \multicolumn{3}{c}{[mmag]} & \colhead{A} & \colhead{R} & }
\startdata
V157 & 184.85275 & 47.23714 &  86.804 & 25.495 & 24.421 & 26.445 & 18 & 12 & 33 & 225 & 163 &  191 & C & C & \\
V158 & 184.64136 & 47.40975 &  88.826 & 25.186 & 24.555 &    \nd & 18 & 15 &\nd & 170 & 138 &  \nd & A & C & \\
V159 & 184.82123 & 47.23845 &  89.470 & 25.096 & 24.370 &    \nd & 16 & 15 &\nd & 205 & 123 &  \nd & B & C & \\
V160 & 184.67403 & 47.36641 &  92.326 & 24.907 & 24.149 &    \nd & 15 & 12 &\nd & 286 & 179 &  \nd & B & C & \\
V161 & 184.84747 & 47.22444 &  92.704 & 25.256 & 24.148 &    \nd & 17 & 10 &\nd & 190 & 106 &  \nd & C & C & \\
V162 & 184.83473 & 47.25617 &  92.821 & 22.957 & 22.064 & 24.506 & 02 & 02 & 07 & 185 & 129 &  315 & A & C & \\
V163 & 184.85008 & 47.21184 &  93.248 & 25.938 & 25.123 &    \nd & 35 & 27 &\nd & 269 & 196 &  \nd & A & C & \\
V164 & 184.86081 & 47.21518 &  93.353 & 24.921 & 24.552 &    \nd & 12 & 15 &\nd & 238 & 182 &  \nd & A & C & \\
V165 & 184.80736 & 47.17226 &  94.460 & 25.054 & 24.602 &    \nd & 12 & 14 &\nd & 161 & 125 &  \nd & A & C & \\
V166 & 184.84497 & 47.25024 &  94.662 & 25.834 & 24.942 &    \nd & 27 & 21 &\nd & 234 & 169 &  \nd & A & C & \\
V167 & 184.73820 & 47.38466 &  95.210 & 25.692 & 24.693 &    \nd & 28 & 17 &\nd & 381 & 204 &  \nd & C & C & \\
V168 & 184.83627 & 47.22005 &  95.212 & 23.407 & 23.045 & 24.215 & 03 & 04 & 06 & 183 & 159 &  359 & A & C & MO028606\\
V169 & 184.67161 & 47.33118 &  95.241 & 25.162 & 24.866 &    \nd & 18 & 22 &\nd & 203 & 209 &  \nd & C & C & \\
V170 & 184.73251 & 47.32790 &  95.324 & 23.944 & 23.701 &    \nd & 12 & 12 &\nd & 150 & 121 &  \nd & A & C & \\
V171 & 184.82220 & 47.18701 &  95.623 & 22.806 & 21.899 & 24.302 & 02 & 01 & 07 & 128 & 100 &  112 & C & C & \\
V172 & 184.81796 & 47.22609 &  95.895 & 25.596 & 24.737 &    \nd & 30 & 20 &\nd & 370 & 211 &  \nd & B & C & \\
V173 & 184.66537 & 47.36115 &  95.932 & 24.176 & 23.637 & 24.875 & 07 & 08 & 11 & 159 & 108 &  265 & A & C & \\
V174 & 184.86279 & 47.16787 &  96.824 & 25.466 & 24.338 &    \nd & 18 & 10 &\nd & 256 & 165 &  \nd & A & C & \\
V175 & 184.80141 & 47.25284 &  97.793 & 25.780 & 24.599 &    \nd & 28 & 15 &\nd & 331 & 200 &  \nd & B & C & \\
V176 & 184.61964 & 47.37447 &  98.284 & 25.091 & 24.332 &    \nd & 17 & 13 &\nd & 174 & 184 &  \nd & C & C & \\
V177 & 184.61723 & 47.33751 & 100.502 & 24.281 & 23.889 & 25.213 & 07 & 07 & 13 & 170 & 173 &  206 & C & C & \\
V178 & 184.70874 & 47.39122 & 101.217 & 23.945 & 23.396 & 25.968 & 07 & 07 & 36 & 167 & 131 &  191 & B & C & \\
V179 & 184.70857 & 47.34545 & 101.461 & 25.025 & 24.600 &    \nd & 22 & 20 &\nd & 234 & 156 &  \nd & A & C & \\
V180 & 184.67085 & 47.36091 & 104.565 & 25.768 & 24.886 &    \nd & 36 & 23 &\nd & 377 & 302 &  \nd & A & C & \\
V181 & 184.66515 & 47.36625 & 105.016 & 25.286 & 24.699 &    \nd & 22 & 19 &\nd & 203 & 200 &  \nd & B & C & \\
V182 & 184.69290 & 47.39526 & 105.526 & 24.807 & 24.327 &    \nd & 13 & 12 &\nd & 218 & 146 &  \nd & A & C & \\
V183 & 184.84706 & 47.17998 & 105.594 & 25.575 & 24.795 &    \nd & 21 & 17 &\nd & 220 & 221 &  \nd & C & C & \\
V184 & 184.83070 & 47.20387 & 106.526 & 23.449 & 22.320 &    \nd & 04 & 02 &\nd & 183 & 125 &  \nd & A & C & \\
V185 & 184.65627 & 47.36112 & 106.700 & 24.507 & 23.550 &    \nd & 11 & 07 &\nd & 143 & 106 &  \nd & A & C & \\
V186 & 184.65965 & 47.34709 & 109.364 & 25.402 & 24.805 &    \nd & 23 & 21 &\nd & 258 & 190 &  \nd & A & C & \\
V187 & 184.70140 & 47.35214 & 109.635 & 24.989 & 24.061 &    \nd & 22 & 13 &\nd & 220 & 165 &  \nd & A & C & \\
V188 & 184.70262 & 47.35945 & 110.489 & 24.584 & 23.590 &    \nd & 18 & 09 &\nd & 112 & 102 &  \nd & A & C & \\
V189 & 184.77916 & 47.20367 & 111.573 & 25.539 & 24.686 &    \nd & 28 & 19 &\nd & 441 & 325 &  \nd & A & C & \\
V190 & 184.73289 & 47.38322 & 111.590 & 25.056 & 23.664 & 26.547 & 19 & 07 & 81 & 260 & 167 &  285 & B & C & \\
V191 & 184.71700 & 47.37902 & 111.750 & 24.231 & 23.922 & 25.239 & 09 & 11 & 21 & 130 & 106 &  174 & A & C & \\
V192 & 184.68226 & 47.33691 & 111.814 & 24.723 & 23.680 &    \nd & 17 & 11 &\nd & 220 & 115 &  \nd & C & C & \\
V193 & 184.82829 & 47.19758 & 113.580 & 25.193 & 24.484 &    \nd & 17 & 15 &\nd & 289 & 234 &  \nd & A & C & \\
V194 & 184.83603 & 47.16718 & 113.615 & 25.079 & 24.652 &    \nd & 14 & 16 &\nd & 185 & 108 &  \nd & B & C & \\
V195 & 184.62598 & 47.35208 & 113.660 & 25.495 & 24.596 &    \nd & 24 & 17 &\nd & 419 & 254 &  \nd & B & C & \\
V196 & 184.71770 & 47.35175 & 113.819 & 23.650 & 22.539 &    \nd & 07 & 03 &\nd & 174 & 109 &  \nd & B & C & \\
V197 & 184.84521 & 47.16626 & 114.106 & 25.438 & 24.471 &    \nd & 21 & 16 &\nd & 289 & 277 &  \nd & B & C & \\
V198 & 184.72798 & 47.36155 & 115.726 & 25.516 & 24.546 &    \nd & 29 & 17 &\nd & 392 & 334 &  \nd & A & C & \\
V199 & 184.81068 & 47.24450 & 118.645 & 25.226 & 24.637 &    \nd & 17 & 17 &\nd & 227 & 200 &  \nd & A & C & \\
V200 & 184.70355 & 47.39479 & 119.050 & 25.560 & 24.674 &    \nd & 25 & 16 &\nd & 207 & 154 &  \nd & A & C & \\
V201 & 184.84435 & 47.17062 & 120.160 & 24.544 & 23.484 & 25.963 & 07 & 05 & 28 & 284 & 144 &  285 & C & C & \\
V202 & 184.71356 & 47.32236 & 120.356 & 22.884 & 22.706 & 23.529 & 05 & 06 & 07 & 271 & 200 &  386 & A & C & MI062769,F55\\
V203 & 184.89284 & 47.18734 & 120.719 & 25.721 & 24.999 &    \nd & 28 & 23 &\nd & 377 & 298 &  \nd & A & C & \\
V204 & 184.82556 & 47.24960 & 121.079 & 25.546 & 24.720 & 26.672 & 22 & 17 & 58 & 372 & 384 &  400 & C & C & \\
V205 & 184.86218 & 47.16662 & 121.893 & 25.711 & 24.762 &    \nd & 23 & 17 &\nd & 242 & 123 &  \nd & C & C & \\
V206 & 184.73367 & 47.35972 & 121.982 & 24.555 & 24.104 &    \nd & 12 & 12 &\nd & 178 & 156 &  \nd & A & C & \\
V207 & 184.81612 & 47.19847 & 124.233 & 25.028 & 24.128 &    \nd & 15 & 10 &\nd & 196 & 165 &  \nd & A & C & \\
V208 & 184.70520 & 47.37148 & 127.054 & 25.059 & 24.133 &    \nd & 18 & 12 &\nd & 242 & 192 &  \nd & A & C & \\
V209 & 184.86861 & 47.23246 & 127.056 & 25.300 & 24.118 &    \nd & 16 & 09 &\nd & 245 & 138 &  \nd & C & C & \\
V210 & 184.82518 & 47.24349 & 127.181 & 25.636 & 24.690 &    \nd & 25 & 17 &\nd & 258 & 181 &  \nd & A & C & \\
V211 & 184.82652 & 47.20412 & 128.765 & 25.314 & 24.525 &    \nd & 19 & 15 &\nd & 236 & 206 &  \nd & A & C & \\
V212 & 184.81662 & 47.19888 & 130.653 & 25.178 & 23.951 &    \nd & 18 & 10 &\nd & 302 & 307 &  \nd & C & C & \\
V213 & 184.83199 & 47.22706 & 135.601 & 25.500 & 24.805 &    \nd & 23 & 21 &\nd & 271 & 148 &  \nd & C & C & \\
V214 & 184.61379 & 47.33673 & 138.177 & 22.252 & 21.447 & 23.895 & 01 & 01 & 04 & 205 & 108 &  171 & C & C & \\
V215 & 184.72481 & 47.32247 & 156.980 & 23.823 & 22.794 &    \nd & 10 & 05 &\nd & 264 & 175 &  \nd & A & C &
\enddata
\tablecomments{The uncertainties in mean magnitude reflect only the statistical component; please refer to Table~\ref{tab:photcal} for systematic uncertainties. Quality flags: A, amplitude ratios; R, P-L residuals.}
\end{deluxetable}

\clearpage

\begin{deluxetable}{llllll}
\tablecaption{Light curve data \label{tab:lcdata}}
\tablewidth{0pc}
\tablehead{
\colhead{ID} & \colhead{MJD} & \colhead{Filter} & \colhead{Mag} & \colhead{$\sigma$} & \colhead{Phase}}
\startdata
C001 & 3053.9805 & g & 25.424 & 57 & 306\\
C001 & 3053.9905 & g & 25.365 & 54 & 307\\
C001 & 3054.0099 & r & 24.642 & 50 & 310\\
C001 & 3054.0185 & r & 24.638 & 60 & 311\\
C001 & 3054.0272 & i & 24.403 & 84 & 313\\ 
C001 & 3054.0358 & i & 24.506 & 74 & 314
\enddata
\tablecomments{This table is available in its entirety in machine-readable form in the online version of the paper. A portion is shown here for guidance regarding its form and content.}
\end{deluxetable}

\begin{figure}[htbp]
\begin{center}
\includegraphics[width=0.49\textwidth]{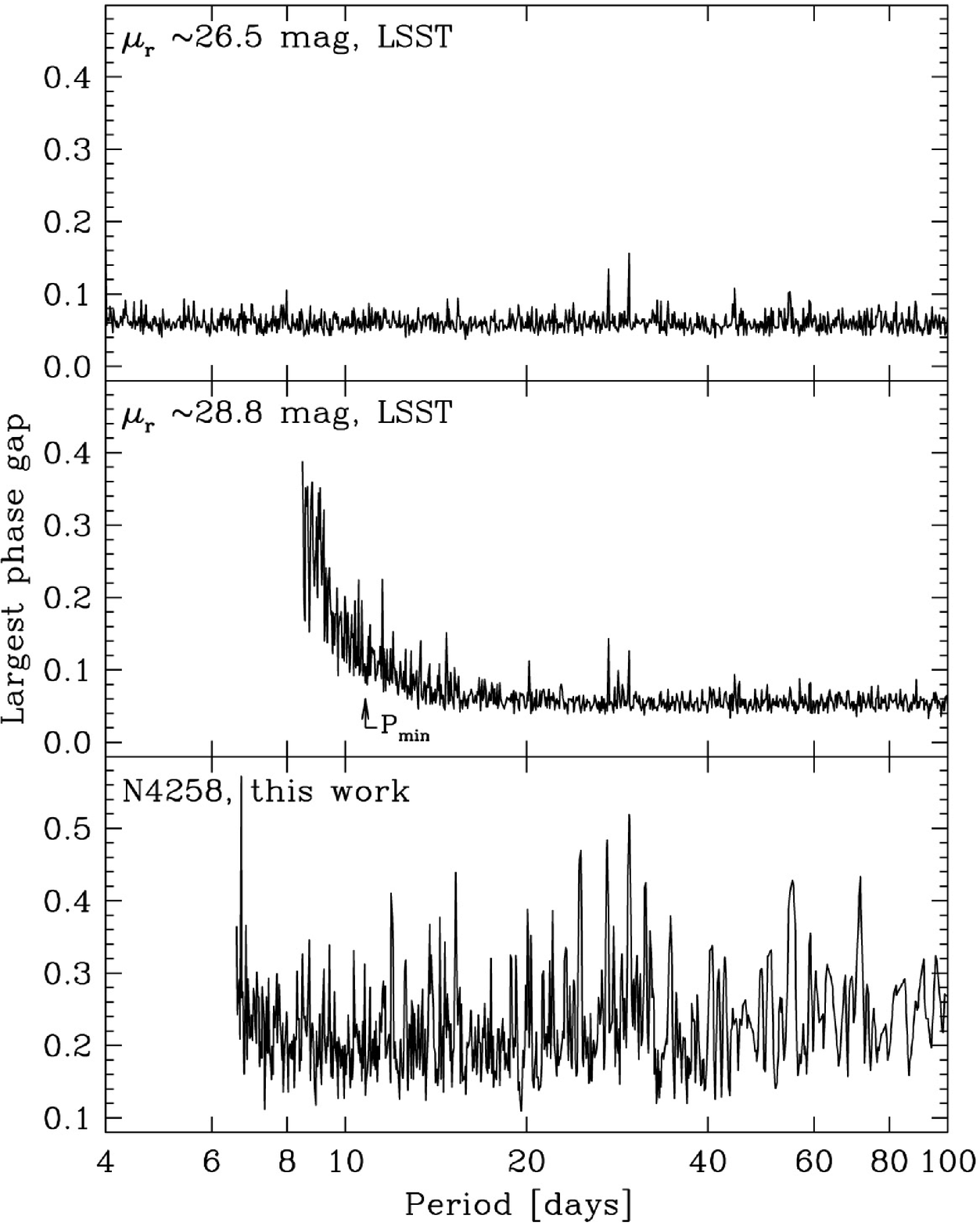}
\end{center}
\caption{Top and middle panels: Maximum phase gap as a function of period in the light curve of Cepheids observed at the expected LSST cadence and {\it gri} magnitude limits, for two galaxies with apparent $r$ distance moduli of 26.5 and 28.8~mag (top and middle, respectively). $P_{min}$ indicates the period below which the maximum phase gap always exceeds the $+3\sigma$ value. Bottom panel: Same as above, but based on the cadence obtained during our survey of \ngal. \label{fig:lsstph}}
\end{figure}

\begin{figure}[htbp]
\begin{center}
\includegraphics[width=0.49\textwidth]{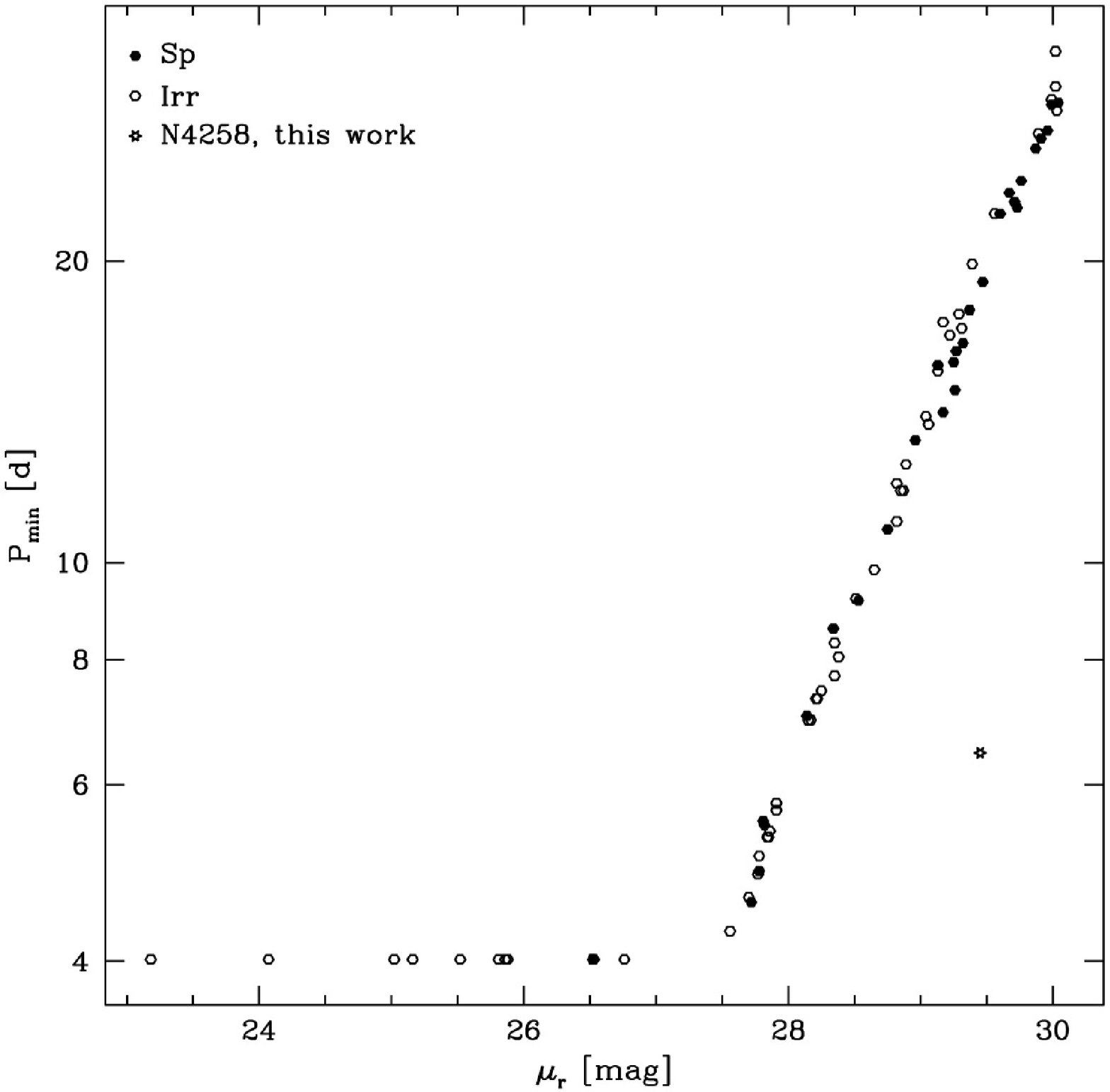}
\end{center}
\vspace{-9pt}
\caption{$P_{min}$ versus distance modulus for the simulated LSST observations. Solid symbols denote spiral galaxies, while open ones represent dwarf galaxies. The star symbol shows the corresponding values for our survey of \ngal. LSST will deliver excellent phase coverage down to $P\!=\!4$d for galaxies with $D\lesssim4.4$~Mpc ($mu\lesssim 27.5$~mag), after which the limiting magnitude will impact the completeness of the P-L relation at the shortest periods.\label{fig:lsstdm}}
\end{figure}

We used the following procedure to calculate the approximate minimum period ($P_{min}$) down to which we would expect complete coverage of the P-L relations of each of the galaxies listed in Table~\ref{tab:lsstgal} in at least one of the {\it gri} bands. We queried the latest realization of the baseline LSST operations over a ten-year period (ops1.1140) and retrieved the Julian Date, seeing, and $5\sigma$ limiting magnitude of the simulated {\it gri} observations, discarding those with image quality worse than $1\arcsec$. We grouped together observations in a given band obtained on the same night into an ``epoch'' with the mean Julian Date and the deepest magnitude limit of an individual image (note that this is a conservative limit, since in a real analysis one would combine all images from a given night to increase the depth of the epoch). The resulting number of epochs per band, average seeing and typical $5\sigma$ limiting magnitudes are those quoted above. Next, we used the EDD distance modulus and value of Galactic extinction for the given galaxy, along with Eqns.~\ref{eqn:plg}-\ref{eqn:pli}, to calculate the faintest apparent magnitude for a Cepheid of a given period, assumed to lie $+2\sigma$ below the mean relation. We combined this information to calculate the shortest Cepheid period that would have complete P-L coverage for each epoch of observation in each band. Once this process was completed, we determined the largest phase gap that would be present in the light curve of a Cepheid of a given period, given the epochs when such a variable could have been detected (above the $5\sigma$ magnitude limit). We carried out this calculation for $10^3$ trial periods equally spaced in logarithmic space for $4 < P < 100$d. Figure~\ref{fig:lsstph} shows the result of this simulation for two of the galaxies, with effective $r$-band distance moduli of 26.5 and 28.6~mag, as well as the phase coverage delivered by our observations of \ngal. Figure~\ref{fig:lsstdm} plots the relation between $P_{min}$ and apparent distance modulus in $r$ for all galaxies listed in the aforementioned Table.

\vspace{3pt}

We found that for galaxies located at $D\lesssim 4.4$~Mpc, the expected LSST cadence and magnitude depth will deliver excellent light curve coverage for all periods of interest. The largest phase gap will typically be $0.058\pm0.01$ or $\sim4\times$ better than our Gemini observations of \ngal, thanks to the significantly larger number of epochs to be obtained. The limiting magnitudes of LSST will result in a increasingly larger value of $P_{min}$ as a function of distance for farther objects, as seen in Fig.~\ref{fig:lsstdm}. Note that this is again a conservative estimate since we were able to determine reliable periods for variables in \ngal\ despite a typical maximum phase gap of 0.2; setting this as the limit for P-L completeness reduces $\log P_{min}$ by $\sim 0.07$~dex, to $P\sim25$d at $D\sim 10$~Mpc.

\section{Summary}
We used GMOS on Gemini North to carry out a synoptic survey of two fields within \ngal\, which resulted in the detection of \nceph\ Cepheid candidates and \nvars\ periodic variables; 262 of these were previously unknown. We derived synthetic P-L relations in the SDSS filters using the Cepheid models of \citet{dicriscienzo13} and found that their absolute calibration yields distance moduli that are in good agreement with the maser distance to this galaxy obtained by \citet{humphreys13}. We investigated the prospects for surveys of extragalactic Population I \& II Cepheids using the expected cadence and depth of LSST and found they bode well for a survey of suitable southern galaxies out to $D\sim 10$~Mpc.

\acknowledgements

SLH and LMM were visiting astronomers at Kitt Peak National Observatory, National Optical Astronomy Observatory, operated by the Association of Universities for Research in Astronomy under cooperative agreement with the National Science Foundation. We acknowledge support by: NASA through the grants HST-GO-09810, -10399, -10802, and -12880; Texas A\&M University through a faculty startup fund; and the Mitchell Institute for Fundamental Physics \& Astronomy at Texas A\&M University. LMM acknowledges initial support for this project by NASA through Hubble Fellowship grant HST-HF-01153 from the Space Telescope Science Institute and by the National Science Foundation through a Goldberg Fellowship from the National Optical Astronomy Observatory. We thank P.~Yoachim for kindly generating $B$-band Cepheid light curve templates.

\ \par

This publication has made use of the following resources:

\begin{itemize}

\item observations obtained at the Gemini Observatory, which is operated by the Association of Universities for Research in Astronomy, Inc., under a cooperative agreement with the NSF on behalf of the Gemini partnership: the National Science Foundation (United States), the Science and Technology Facilities Council (United Kingdom), the National Research Council (Canada), CONICYT (Chile), the Australian Research Council (Australia), Minist\'{e}rio da Ci\^{e}ncia, Tecnologia e Inova\c{c}\~{a}o (Brazil) and Ministerio de Ciencia, Tecnolog\'{i}a e Innovaci\'{o}n Productiva (Argentina).

\item observations obtained at the WIYN Observatory, which is a joint facility of the University of Wisconsin-Madison, Indiana University, Yale University and the National Optical Astronomy Observatory.

\item data products from the Sloan Digital Sky Survey (SDSS and SDSS-II), funded by the Alfred P. Sloan Foundation, the Participating Institutions, the National Science Foundation, the U.S. Department of Energy, the National Aeronautics and Space Administration, the Japanese Monbukagakusho, the Max Planck Society, and the Higher Education Funding Council for England. The SDSS Web Site is http://www.sdss.org/ . The SDSS is managed by the Astrophysical Research Consortium for the Participating Institutions. The Participating Institutions are the American Museum of Natural History, Astrophysical Institute Potsdam, University of Basel, University of Cambridge, Case Western Reserve University, University of Chicago, Drexel University, Fermilab, the Institute for Advanced Study, the Japan Participation Group, Johns Hopkins University, the Joint Institute for Nuclear Astrophysics, the Kavli Institute for Particle Astrophysics and Cosmology, the Korean Scientist Group, the Chinese Academy of Sciences (LAMOST), Los Alamos National Laboratory, the Max-Planck-Institute for Astronomy (MPIA), the Max-Planck-Institute for Astrophysics (MPA), New Mexico State University, Ohio State University, University of Pittsburgh, University of Portsmouth, Princeton University, the United States Naval Observatory, and the University of Washington.

\item Montage, funded by the National Aeronautics and Space Administration's Earth Science Technology Office, Computational Technologies Project, under Cooperative Agreement Number NCC5-626 between NASA and the California Institute of Technology. The code is maintained by the NASA/IPAC Infrared Science Archive.

\item NASA's Astrophysics Data System at the Harvard-Smithsonian Center for Astrophysics.

\item the NASA/IPAC Extragalactic Database (NED) which is operated by the Jet Propulsion Laboratory, California Institute of Technology, under contract with the National Aeronautics and Space Administration.

\item the Extragalactic Distance Database (EDD), with support for the development of its content provided by the National Science Foundation under Grant No. AST09-08846.
\end{itemize}

{\it Facilities:} \facility{Gemini:Gillett},\facility{WIYN}

\bibliography{hoffmann}{}

\begin{thebibliography}{51}
\expandafter\ifx\csname natexlab\endcsname\relax\def\natexlab#1{#1}\fi

\bibitem[{{Abazajian} {et~al.}(2009){Abazajian}, {Adelman-McCarthy},
  {Ag{\"u}eros}, {Allam}, {Allende Prieto}, {An}, {Anderson}, {Anderson},
  {Annis}, {Bahcall}, \& et~al.}]{abazajian09}
{Abazajian}, K.~N., {Adelman-McCarthy}, J.~K., {Ag{\"u}eros}, M.~A., {Allam},
  S.~S., {Allende Prieto}, C., {An}, D., {Anderson}, K.~S.~J., {Anderson},
  S.~F., {Annis}, J., {Bahcall}, N.~A., \& et~al. 2009, \apjs, 182, 543

\bibitem[{{Anderson} {et~al.}(2014){Anderson}, {Aubourg}, {Bailey}, {Beutler},
  {Bhardwaj}, {Blanton}, {Bolton}, {Brinkmann}, {Brownstein}, {Burden}, \&
  et~al.}]{anderson14}
{Anderson}, L., {Aubourg}, {\'E}., {Bailey}, S., {Beutler}, F., {Bhardwaj}, V.,
  {Blanton}, M., {Bolton}, A.~S., {Brinkmann}, J., {Brownstein}, J.~R.,
  {Burden}, A., \& et~al. 2014, \mnras, 441, 24

\bibitem[{{Argon} {et~al.}(2007){Argon}, {Greenhill}, {Reid}, {Moran}, \&
  {Humphreys}}]{argon07}
{Argon}, A.~L., {Greenhill}, L.~J., {Reid}, M.~J., {Moran}, J.~M., \&
  {Humphreys}, E.~M.~L. 2007, \apj, 659, 1040

\bibitem[{{Barnes} {et~al.}(1999){Barnes}, {Ivans}, {Martin}, {Froning}, \&
  {Moffett}}]{barnes99}
{Barnes}, III, T.~G., {Ivans}, I.~I., {Martin}, J.~R., {Froning}, C.~S., \&
  {Moffett}, T.~J. 1999, \pasp, 111, 812

\bibitem[{{Bonanos} \& {Stanek}(2003)}]{bonanos03}
{Bonanos}, A.~Z. \& {Stanek}, K.~Z. 2003, \apjl, 591, L111

\bibitem[{{Bonanos} {et~al.}(2006){Bonanos}, {Stanek}, {Kudritzki}, {Macri},
  {Sasselov}, {Kaluzny}, {Stetson}, {Bersier}, {Bresolin}, {Matheson},
  {Mochejska}, {Przybilla}, {Szentgyorgyi}, {Tonry}, \& {Torres}}]{bonanos06}
{Bonanos}, A.~Z., {Stanek}, K.~Z., {Kudritzki}, R.~P., {Macri}, L.~M.,
  {Sasselov}, D.~D., {Kaluzny}, J., {Stetson}, P.~B., {Bersier}, D.,
  {Bresolin}, F., {Matheson}, T., {Mochejska}, B.~J., {Przybilla}, N.,
  {Szentgyorgyi}, A.~H., {Tonry}, J., \& {Torres}, G. 2006, \apj, 652, 313

\bibitem[{{Bresolin}(2011)}]{bresolin11}
{Bresolin}, F. 2011, \apj, 729, 56

\bibitem[{{Castelli} \& {Kurucz}(2003)}]{castelli03}
{Castelli}, F. \& {Kurucz}, R.~L. 2003, in IAU Symposium, Vol. 210, Modelling
  of Stellar Atmospheres, ed. N.~{Piskunov}, W.~W. {Weiss}, \& D.~F. {Gray},
  20P

\bibitem[{{Davies} {et~al.}(1997){Davies}, {Allington-Smith}, {Bettess},
  {Chadwick}, {Content}, {Dodsworth}, {Haynes}, {Lee}, {Lewis}, {Webster},
  {Atad}, {Beard}, {Ellis}, {Hastings}, {Williams}, {Bond}, {Crampton},
  {Davidge}, {Fletcher}, {Leckie}, {Morbey}, {Murowinski}, {Roberts},
  {Saddlemyer}, {Sebesta}, {Stilburn}, \& {Szeto}}]{davies97}
{Davies}, R.~L., {Allington-Smith}, J.~R., {Bettess}, P., {Chadwick}, E.,
  {Content}, R., {Dodsworth}, G.~N., {Haynes}, R., {Lee}, D., {Lewis}, I.~J.,
  {Webster}, J., {Atad}, E., {Beard}, S.~M., {Ellis}, M., {Hastings}, P.~R.,
  {Williams}, P.~R., {Bond}, T., {Crampton}, D., {Davidge}, T.~J., {Fletcher},
  M., {Leckie}, B., {Morbey}, C.~L., {Murowinski}, R.~G., {Roberts}, S.,
  {Saddlemyer}, L.~K., {Sebesta}, J., {Stilburn}, J.~R., \& {Szeto}, K. 1997,
  in Society of Photo-Optical Instrumentation Engineers (SPIE) Conference
  Series, Vol. 2871, Society of Photo-Optical Instrumentation Engineers (SPIE)
  Conference Series, ed. A.~L. {Ardeberg}, 1099--1106

\bibitem[{{Di Criscienzo} {et~al.}(2013){Di Criscienzo}, {Marconi}, {Musella},
  {Cignoni}, \& {Ripepi}}]{dicriscienzo13}
{Di Criscienzo}, M., {Marconi}, M., {Musella}, I., {Cignoni}, M., \& {Ripepi},
  V. 2013, \mnras, 428, 212

\bibitem[{{Dvorkin} {et~al.}(2014){Dvorkin}, {Wyman}, {Rudd}, \&
  {Hu}}]{dvorkin14}
{Dvorkin}, C., {Wyman}, M., {Rudd}, D.~H., \& {Hu}, W. 2014, \prd, 90, 083503

\bibitem[{{Fausnaugh} {et~al.}(2014){Fausnaugh}, {Kochanek}, {Gerke}, {Macri},
  {Riess}, \& {Stanek}}]{fausnaugh14}
{Fausnaugh}, M.~M., {Kochanek}, C.~S., {Gerke}, J.~R., {Macri}, L.~M., {Riess},
  A.~G., \& {Stanek}, K.~Z. 2014, ArXiv e-prints

\bibitem[{{Ferrarese} {et~al.}(2007){Ferrarese}, {Mould}, {Stetson}, {Tonry},
  {Blakeslee}, \& {Ajhar}}]{ferrarese07}
{Ferrarese}, L., {Mould}, J.~R., {Stetson}, P.~B., {Tonry}, J.~L., {Blakeslee},
  J.~P., \& {Ajhar}, E.~A. 2007, \apj, 654, 186

\bibitem[{{Fitzpatrick}(1999)}]{fitzpatrick99}
{Fitzpatrick}, E.~L. 1999, \pasp, 111, 63

\bibitem[{{Freedman} {et~al.}(1985){Freedman}, {Grieve}, \&
  {Madore}}]{freedman85}
{Freedman}, W.~L., {Grieve}, G.~R., \& {Madore}, B.~F. 1985, \apjs, 59, 311

\bibitem[{{Fukugita} {et~al.}(1996){Fukugita}, {Ichikawa}, {Gunn}, {Doi},
  {Shimasaku}, \& {Schneider}}]{fukugita96}
{Fukugita}, M., {Ichikawa}, T., {Gunn}, J.~E., {Doi}, M., {Shimasaku}, K., \&
  {Schneider}, D.~P. 1996, \aj, 111, 1748

\bibitem[{{Gerke} {et~al.}(2011){Gerke}, {Kochanek}, {Prieto}, {Stanek}, \&
  {Macri}}]{gerke11}
{Gerke}, J.~R., {Kochanek}, C.~S., {Prieto}, J.~L., {Stanek}, K.~Z., \&
  {Macri}, L.~M. 2011, \apj, 743, 176

\bibitem[{{Hartman} {et~al.}(2006){Hartman}, {Bersier}, {Stanek}, {Beaulieu},
  {Kaluzny}, {Marquette}, {Stetson}, \& {Schwarzenberg-Czerny}}]{hartman06}
{Hartman}, J.~D., {Bersier}, D., {Stanek}, K.~Z., {Beaulieu}, J.-P., {Kaluzny},
  J., {Marquette}, J.-B., {Stetson}, P.~B., \& {Schwarzenberg-Czerny}, A. 2006,
  \mnras, 371, 1405

\bibitem[{{Herrnstein} {et~al.}(1999){Herrnstein}, {Moran}, {Greenhill},
  {Diamond}, {Inoue}, {Nakai}, {Miyoshi}, {Henkel}, \& {Riess}}]{herrnstein99}
{Herrnstein}, J.~R., {Moran}, J.~M., {Greenhill}, L.~J., {Diamond}, P.~J.,
  {Inoue}, M., {Nakai}, N., {Miyoshi}, M., {Henkel}, C., \& {Riess}, A. 1999,
  \nat, 400, 539

\bibitem[{{Humphreys} {et~al.}(2008){Humphreys}, {Reid}, {Greenhill}, {Moran},
  \& {Argon}}]{humphreys08}
{Humphreys}, E.~M.~L., {Reid}, M.~J., {Greenhill}, L.~J., {Moran}, J.~M., \&
  {Argon}, A.~L. 2008, \apj, 672, 800

\bibitem[{{Humphreys} {et~al.}(2013){Humphreys}, {Reid}, {Moran}, {Greenhill},
  \& {Argon}}]{humphreys13}
{Humphreys}, E.~M.~L., {Reid}, M.~J., {Moran}, J.~M., {Greenhill}, L.~J., \&
  {Argon}, A.~L. 2013, \apj, 775, 13

\bibitem[{{J{\o}rgensen}(2009)}]{jorgensen09}
{J{\o}rgensen}, I. 2009, {Publications of the Astronomical Society of
  Australia}, 26, 17

\bibitem[{{Kodric} {et~al.}(2013){Kodric}, {Riffeser}, {Hopp}, {Seitz},
  {Koppenhoefer}, {Bender}, {Goessl}, {Snigula}, {Lee}, {Ngeow}, {Chambers},
  {Magnier}, {Price}, {Burgett}, {Hodapp}, {Kaiser}, \& {Kudritzki}}]{kodric13}
{Kodric}, M., {Riffeser}, A., {Hopp}, U., {Seitz}, S., {Koppenhoefer}, J.,
  {Bender}, R., {Goessl}, C., {Snigula}, J., {Lee}, C.-H., {Ngeow}, C.-C.,
  {Chambers}, K.~C., {Magnier}, E.~A., {Price}, P.~A., {Burgett}, W.~S.,
  {Hodapp}, K.~W., {Kaiser}, N., \& {Kudritzki}, R.-P. 2013, \aj, 145, 106

\bibitem[{{Kodric} {et~al.}(2014){Kodric}, {Riffeser}, {Seitz}, {Snigula},
  {Hopp}, {Lee}, {Goessl}, {Koppenhoefer}, {Bender}, \& {Gieren}}]{kodric14}
{Kodric}, M., {Riffeser}, A., {Seitz}, S., {Snigula}, J., {Hopp}, U., {Lee},
  C.-H., {Goessl}, C., {Koppenhoefer}, J., {Bender}, R., \& {Gieren}, W. 2014,
  ArXiv e-prints

\bibitem[{{Macri} {et~al.}(2014){Macri}, {Ngeow}, {Kanbur}, {Mahzooni}, \&
  {Smitka}}]{macri14}
{Macri}, L.~M., {Ngeow}, C.-C., {Kanbur}, S., {Mahzooni}, S., \& {Smitka}, M.
  2014, {submitted to \aj}

\bibitem[{{Macri} {et~al.}(2006){Macri}, {Stanek}, {Bersier}, {Greenhill}, \&
  {Reid}}]{macri06}
{Macri}, L.~M., {Stanek}, K.~Z., {Bersier}, D., {Greenhill}, L.~J., \& {Reid},
  M.~J. 2006, \apj, 652, 1133

\bibitem[{{Martin} {et~al.}(1979){Martin}, {Warren}, \& {Feast}}]{martin79}
{Martin}, W.~L., {Warren}, P.~R., \& {Feast}, M.~W. 1979, \mnras, 188, 139

\bibitem[{{Miyoshi} {et~al.}(1995){Miyoshi}, {Moran}, {Herrnstein},
  {Greenhill}, {Nakai}, {Diamond}, \& {Inoue}}]{miyoshi95}
{Miyoshi}, M., {Moran}, J., {Herrnstein}, J., {Greenhill}, L., {Nakai}, N.,
  {Diamond}, P., \& {Inoue}, M. 1995, \nat, 373, 127

\bibitem[{{Ngeow} \& {Kanbur}(2007)}]{ngeow07}
{Ngeow}, C. \& {Kanbur}, S.~M. 2007, \apj, 667, 35

\bibitem[{{Ngeow} \& {Kanbur}(2006)}]{ngeow06}
{Ngeow}, C.-C. \& {Kanbur}, S.~M. 2006, \mnras, 369, 723

\bibitem[{{Pietrzy{\'n}ski} {et~al.}(2013){Pietrzy{\'n}ski}, {Graczyk},
  {Gieren}, {Thompson}, {Pilecki}, {Udalski}, {Soszy{\'n}ski}, {Koz{\l}owski},
  {Konorski}, {Suchomska}, {Bono}, {Moroni}, {Villanova}, {Nardetto},
  {Bresolin}, {Kudritzki}, {Storm}, {Gallenne}, {Smolec}, {Minniti}, {Kubiak},
  {Szyma{\'n}ski}, {Poleski}, {Wyrzykowski}, {Ulaczyk}, {Pietrukowicz},
  {G{\'o}rski}, \& {Karczmarek}}]{pietrzynski13}
{Pietrzy{\'n}ski}, G., {Graczyk}, D., {Gieren}, W., {Thompson}, I.~B.,
  {Pilecki}, B., {Udalski}, A., {Soszy{\'n}ski}, I., {Koz{\l}owski}, S.,
  {Konorski}, P., {Suchomska}, K., {Bono}, G., {Moroni}, P.~G.~P., {Villanova},
  S., {Nardetto}, N., {Bresolin}, F., {Kudritzki}, R.~P., {Storm}, J.,
  {Gallenne}, A., {Smolec}, R., {Minniti}, D., {Kubiak}, M., {Szyma{\'n}ski},
  M.~K., {Poleski}, R., {Wyrzykowski}, {\L}., {Ulaczyk}, K., {Pietrukowicz},
  P., {G{\'o}rski}, M., \& {Karczmarek}, P. 2013, \nat, 495, 76

\bibitem[{{Planck Collaboration} {et~al.}(2013){Planck Collaboration}, {Ade},
  {Aghanim}, {Armitage-Caplan}, {Arnaud}, {Ashdown}, {Atrio-Barandela},
  {Aumont}, {Baccigalupi}, {Banday}, \& et~al.}]{planck13}
{Planck Collaboration}, {Ade}, P.~A.~R., {Aghanim}, N., {Armitage-Caplan}, C.,
  {Arnaud}, M., {Ashdown}, M., {Atrio-Barandela}, F., {Aumont}, J.,
  {Baccigalupi}, C., {Banday}, A.~J., \& et~al. 2013, ArXiv e-prints

\bibitem[{{Rejkuba}(2004)}]{rejkuba04}
{Rejkuba}, M. 2004, \aap, 413, 903

\bibitem[{{Ribas} {et~al.}(2005){Ribas}, {Jordi}, {Vilardell}, {Fitzpatrick},
  {Hilditch}, \& {Guinan}}]{ribas05}
{Ribas}, I., {Jordi}, C., {Vilardell}, F., {Fitzpatrick}, E.~L., {Hilditch},
  R.~W., \& {Guinan}, E.~F. 2005, \apjl, 635, L37

\bibitem[{{Riess} {et~al.}(2011){Riess}, {Macri}, {Casertano}, {Lampeitl},
  {Ferguson}, {Filippenko}, {Jha}, {Li}, \& {Chornock}}]{riess11}
{Riess}, A.~G., {Macri}, L., {Casertano}, S., {Lampeitl}, H., {Ferguson},
  H.~C., {Filippenko}, A.~V., {Jha}, S.~W., {Li}, W., \& {Chornock}, R. 2011,
  \apj, 730, 119

\bibitem[{{Schlafly} \& {Finkbeiner}(2011)}]{schlafly11}
{Schlafly}, E.~F. \& {Finkbeiner}, D.~P. 2011, \apj, 737, 103

\bibitem[{{Sebo} {et~al.}(2002){Sebo}, {Rawson}, {Mould}, {Madore}, {Putman},
  {Graham}, {Freedman}, {Gibson}, \& {Germany}}]{sebo02}
{Sebo}, K.~M., {Rawson}, D., {Mould}, J., {Madore}, B.~F., {Putman}, M.~E.,
  {Graham}, J.~A., {Freedman}, W.~L., {Gibson}, B.~K., \& {Germany}, L.~M.
  2002, \apjs, 142, 71

\bibitem[{{Soszynski} {et~al.}(2008){Soszynski}, {Poleski}, {Udalski},
  {Szymanski}, {Kubiak}, {Pietrzynski}, {Wyrzykowski}, {Szewczyk}, \&
  {Ulaczyk}}]{soszynski08}
{Soszynski}, I., {Poleski}, R., {Udalski}, A., {Szymanski}, M.~K., {Kubiak},
  M., {Pietrzynski}, G., {Wyrzykowski}, L., {Szewczyk}, O., \& {Ulaczyk}, K.
  2008, \actaa, 58, 163

\bibitem[{{Stetson}(1987)}]{stetson87}
{Stetson}, P.~B. 1987, \pasp, 99, 191

\bibitem[{{Stetson}(1993)}]{stetson93}
{Stetson}, P.~B. 1993, in IAU Colloq. 136: Stellar Photometry - Current
  Techniques and Future Developments, ed. C.~J. {Butler} \& I.~{Elliott}, 291

\bibitem[{{Stetson}(1994)}]{stetson94}
---. 1994, \pasp, 106, 250

\bibitem[{{Stetson}(1996)}]{stetson96}
---. 1996, \pasp, 108, 851

\bibitem[{{Tanvir} \& {Boyle}(1999)}]{tanvir99}
{Tanvir}, N.~R. \& {Boyle}, A. 1999, \mnras, 304, 957

\bibitem[{{Thim} {et~al.}(2003){Thim}, {Tammann}, {Saha}, {Dolphin}, {Sandage},
  {Tolstoy}, \& {Labhardt}}]{thim03}
{Thim}, F., {Tammann}, G.~A., {Saha}, A., {Dolphin}, A., {Sandage}, A.,
  {Tolstoy}, E., \& {Labhardt}, L. 2003, \apj, 590, 256

\bibitem[{{Tully} {et~al.}(2013){Tully}, {Courtois}, {Dolphin}, {Fisher},
  {H{\'e}raudeau}, {Jacobs}, {Karachentsev}, {Makarov}, {Makarova},
  {Mitronova}, {Rizzi}, {Shaya}, {Sorce}, \& {Wu}}]{tully13}
{Tully}, R.~B., {Courtois}, H.~M., {Dolphin}, A.~E., {Fisher}, J.~R.,
  {H{\'e}raudeau}, P., {Jacobs}, B.~A., {Karachentsev}, I.~D., {Makarov}, D.,
  {Makarova}, L., {Mitronova}, S., {Rizzi}, L., {Shaya}, E.~J., {Sorce}, J.~G.,
  \& {Wu}, P.-F. 2013, \aj, 146, 86

\bibitem[{{Tully} {et~al.}(2009){Tully}, {Rizzi}, {Shaya}, {Courtois},
  {Makarov}, \& {Jacobs}}]{tully09}
{Tully}, R.~B., {Rizzi}, L., {Shaya}, E.~J., {Courtois}, H.~M., {Makarov},
  D.~I., \& {Jacobs}, B.~A. 2009, \aj, 138, 323

\bibitem[{{Udalski} {et~al.}(1999){Udalski}, {Soszynski}, {Szymanski},
  {Kubiak}, {Pietrzynski}, {Wozniak}, \& {Zebrun}}]{udalski99}
{Udalski}, A., {Soszynski}, I., {Szymanski}, M., {Kubiak}, M., {Pietrzynski},
  G., {Wozniak}, P., \& {Zebrun}, K. 1999, \actaa, 49, 223

\bibitem[{{Ulaczyk} {et~al.}(2013){Ulaczyk}, {Szyma{\'n}ski}, {Udalski},
  {Kubiak}, {Pietrzy{\'n}ski}, {Soszy{\'n}ski}, {Wyrzykowski}, {Poleski},
  {Gieren}, {Walker}, \& {Garcia-Varela}}]{ulaczyk13}
{Ulaczyk}, K., {Szyma{\'n}ski}, M.~K., {Udalski}, A., {Kubiak}, M.,
  {Pietrzy{\'n}ski}, G., {Soszy{\'n}ski}, I., {Wyrzykowski}, {\L}., {Poleski},
  R., {Gieren}, W., {Walker}, A.~R., \& {Garcia-Varela}, A. 2013, \actaa, 63,
  159

\bibitem[{{Vilardell} {et~al.}(2010){Vilardell}, {Ribas}, {Jordi},
  {Fitzpatrick}, \& {Guinan}}]{vilardell10}
{Vilardell}, F., {Ribas}, I., {Jordi}, C., {Fitzpatrick}, E.~L., \& {Guinan},
  E.~F. 2010, \aap, 509, A70

\bibitem[{{Weinberg} {et~al.}(2013){Weinberg}, {Mortonson}, {Eisenstein},
  {Hirata}, {Riess}, \& {Rozo}}]{weinberg13}
{Weinberg}, D.~H., {Mortonson}, M.~J., {Eisenstein}, D.~J., {Hirata}, C.,
  {Riess}, A.~G., \& {Rozo}, E. 2013, \physrep, 530, 87

\bibitem[{{Yoachim} {et~al.}(2009){Yoachim}, {McCommas}, {Dalcanton}, \&
  {Williams}}]{yoachim09}
{Yoachim}, P., {McCommas}, L.~P., {Dalcanton}, J.~J., \& {Williams}, B.~F.
  2009, \aj, 137, 4697

\end{thebibliography}
\bibliographystyle{apj}

\clearpage

\begin{deluxetable}{rlllccrrl}
\tablecaption{Galaxies suitable for Cepheid searches with LSST \label{tab:lsstgal}}
\tablewidth{\textwidth}
\tablehead{
\colhead{PGC} & \colhead{RA} & \colhead{Dec}      & \colhead{$\mu_0$} & \colhead{C?} & \colhead{T?} & \colhead{Morph.} & \colhead{$i$}   & \colhead{Common}\\
\colhead{}    & \multicolumn{2}{c}{(J2000)}       & \colhead{}        & \colhead{}   & \colhead{}   & \colhead{type}   & \colhead{}      & \colhead{name}\\
\colhead{}    & \colhead{[hms]} & \colhead{[dms]} & \colhead{[mag]}   & \colhead{}   & \colhead{}   & \colhead{}       & \colhead{[deg]} & \colhead{}}
\startdata 
          143 &  00:01:58.2 & -15:27:39           & $24.92\pm0.05$    & \checkmark      & \checkmark      &       10         &                 & WLM                  \\
          621 &  00:08:13.5 & -34:34:43           & $27.53\pm0.10$    &                 & \checkmark      &       10         &                 & ESO349-031           \\
          701 &  00:09:56.3 & -24:57:50           & $29.42\pm0.09$    &                 & \checkmark      &        5         &     79          & N24                  \\
          930 &  00:14:03.9 & -23:10:56           & $29.11\pm0.10$    &                 & \checkmark      &        8         &     43          & NGC45                \\
         1014 &  00:14:53.6 & -39:11:48           & $26.49\pm0.06$    & \checkmark      & \checkmark      &        9         &     74          & N55                  \\
         2142 &  00:35:46.6 & -25:22:27           & $29.84\pm0.20$    &                 &                 &        9         &     39          & I1558                \\
         2578 &  00:43:03.6 & -22:14:51           & $28.46\pm0.10$    &                 & \checkmark      &       10         &                 & DDO226               \\
         2758 &  00:47:08.6 & -20:45:38           & $27.73\pm0.06$    & \checkmark      & \checkmark      &        7         &     73          & N247                 \\
         2789 &  00:47:33.1 & -25:17:18           & $27.76\pm0.08$    &                 & \checkmark      &        5         &     76          & N253                 \\
         2881 &  00:49:21.1 & -18:04:31           & $27.71\pm0.08$    &                 & \checkmark      &        9         &     45          & ESO540-030           \\
         2902 &  00:49:49.7 & -21:00:47           & $27.65\pm0.08$    &                 & \checkmark      &       10         &                 & DDO6                 \\
         2933 &  00:50:24.6 & -19:54:23           & $27.78\pm0.08$    &                 & \checkmark      &       10         &                 & ESO540-032           \\
         3238 &  00:54:53.5 & -37:41:04           & $26.48\pm0.06$    & \checkmark      & \checkmark      &        7         &     48          & N300                 \\
         5896 &  01:35:05.1 & -41:26:12           & $27.73\pm0.09$    &                 & \checkmark      &        9         &     72          & N625                 \\
         6430 &  01:45:03.9 & -43:35:55           & $28.30\pm0.10$    &                 & \checkmark      &       10         &                 & ESO245-005           \\
         6830 &  01:51:06.3 & -44:26:41           & $23.13\pm0.06$    &                 & \checkmark      &       10         &                 & Phoenix              \\
        11211 &  02:58:04.1 & -49:22:56           & $28.90\pm0.10$    &                 & \checkmark      &        8         &     66          & ESO199-007           \\
        11812 &  03:09:38.3 & -41:01:55           & $29.97\pm0.20$    &                 &                 &        9         &     60          & ES300-014            \\
        12460 &  03:20:07.0 & -52:11:09           & $28.59\pm0.09$    &                 & \checkmark      &        9         &     73          & N1311                \\
        13163 &  03:33:12.6 & -50:24:51           & $29.10\pm0.09$    &                 & \checkmark      &        9         &     78          & I1959                \\
        13368 &  03:37:28.3 & -24:30:05           & $29.90\pm0.20$    &                 &                 &        6         &     54          & N1385                \\
        13794 &  03:45:54.9 & -36:21:25           & $29.67\pm0.20$    &                 &                 &        7         &     71          & N1437B               \\
        14475 &  04:06:48.9 & -21:10:41           & $29.59\pm0.20$    &                 &                 &        8         &     62          & N1518                \\
        14897 &  04:20:00.4 & -54:56:16           & $29.10\pm0.20$    &                 &                 &        4         &     46          & N1566                \\
        16120 &  04:49:55.6 & -31:57:56           & $29.93\pm0.20$    &                 &                 &       10         &                 & N1679                \\
        16389 &  04:56:58.7 & -42:48:02           & $29.21\pm0.10$    &                 & \checkmark      &        8         &     41          & ESO252-001           \\
        16517 &  04:59:58.1 & -26:01:30           & $29.92\pm0.20$    &                 &                 &        7         &     60          & N1744                \\
        16779 &  05:07:42.3 & -37:30:47           & $29.79\pm0.20$    &                 &                 &        1         &     46          & N1808                \\
        17302 &  05:27:05.8 & -20:40:40           & $29.13\pm0.10$    &                 & \checkmark      &        4         &     48          & ESO553-046           \\
        18431 &  06:07:19.8 & -34:12:16           & $29.92\pm0.10$    &                 & \checkmark      &       10         &                 & AM0605-341           \\
        18731 &  06:15:54.3 & -57:43:32           & $28.92\pm0.10$    &                 & \checkmark      &       10         &                 & ESO121-020           \\
        19041 &  06:26:17.5 & -26:15:57           & $29.00\pm0.10$    &                 & \checkmark      &       10         &                 & ESO489-056           \\
        19337 &  06:37:57.1 & -26:00:01           & $29.01\pm0.10$    &                 & \checkmark      &       10         &                 & ESO490-017           \\
        26259 &  09:17:52.9 & -22:21:17           & $29.71\pm0.20$    &                 &                 &        5         &     44          & N2835                \\
        29128 &  10:03:06.9 & -26:09:34           & $25.69\pm0.06$    & \checkmark      & \checkmark      &        9         &     78          & N3109                \\
        29194 &  10:04:04.0 & -27:19:55           & $25.64\pm0.08$    &                 & \checkmark      &       10         &                 & Antlia               \\
        29653 &  10:11:00.8 & -04:41:34           & $25.69\pm0.06$    & \checkmark      & \checkmark      &       10         &                 & SextansA             \\
       490287 &  10:57:30.0 & -48:11:02           & $28.69\pm0.10$    &                 & \checkmark      &       10         &                 & ESO215-009           \\
        34554 &  11:18:16.5 & -32:48:50           & $29.14\pm0.06$    & \checkmark      & \checkmark      &        7         &     60          & N3621                \\
        36014 &  11:37:53.2 & -39:13:14           & $28.89\pm0.10$    &                 & \checkmark      &       10         &                 & ESO320-014           \\
        37369 &  11:54:43.2 & -33:33:32           & $28.68\pm0.10$    &                 & \checkmark      &       10         &                 & ESO379-007           \\
        39032 &  12:13:49.7 & -38:13:52           & $27.58\pm0.08$    &                 & \checkmark      &       10         &                 & ESO321-014           \\
        42936 &  12:44:42.5 & -35:57:60           & $28.68\pm0.10$    &                 & \checkmark      &       10         &                 & ESO381-018           \\
        43048 &  12:46:00.4 & -33:50:17           & $28.69\pm0.10$    &                 & \checkmark      &       10         &                 & ESO381-020           \\
        43978 &  12:54:53.6 & -28:20:27           & $28.88\pm0.10$    &                 & \checkmark      &       10         &                 & ESO443-009           \\
        45104 &  13:03:33.2 & -46:35:13           & $27.49\pm0.10$    &                 & \checkmark      &       10         &                 & ESO269-037           \\
        45279 &  13:05:27.3 & -49:28:05           & $27.85\pm0.08$    &                 & \checkmark      &        6         &     77          & N4945                \\
        45717 &  13:10:32.9 & -46:59:31           & $27.87\pm0.10$    &                 & \checkmark      &       10         &                 & ESO269-058           \\
        46663 &  13:21:47.1 & -45:03:45           & $27.99\pm0.10$    &                 & \checkmark      &       10         &                 & KK196                \\
        46938 &  13:25:18.5 & -21:08:03           & $29.06\pm0.20$    &                 &                 &        3         &     47          & N5134                \\
        46957 &  13:25:28.1 & -43:01:05           & $27.82\pm0.06$    & \checkmark      & \checkmark      &       -2         &                 & N5128                \\
        47073 &  13:26:44.4 & -30:21:45           & $28.57\pm0.10$    &                 & \checkmark      &        7         &     66          & IC4247               \\
        47171 &  13:27:38.4 & -41:28:42           & $27.89\pm0.10$    &                 & \checkmark      &       10         &                 & ESO324-24            \\
        48029 &  13:36:30.8 & -29:14:07           & $28.67\pm0.10$    &                 & \checkmark      &       10         &                 & ESO444-78            \\
        48082 &  13:37:00.9 & -29:51:56           & $28.34\pm0.07$    & \checkmark      & \checkmark      &        5         &     32          & M83                  \\
        48334 &  13:39:56.0 & -31:38:24           & $27.75\pm0.06$    & \checkmark      & \checkmark      &        9         &     64          & NGC5253              \\
        48368 &  13:40:18.3 & -28:53:39           & $28.19\pm0.10$    &                 & \checkmark      &       10         &                 & IC4316               \\
        48467 &  13:41:36.7 & -29:54:46           & $28.24\pm0.10$    &                 & \checkmark      &       10         &                 & N5264                \\
        48738 &  13:45:01.0 & -41:51:35           & $27.66\pm0.10$    &                 & \checkmark      &       10         &                 & ESO325-11            \\
        49050 &  13:49:17.5 & -36:03:48           & $27.52\pm0.10$    &                 & \checkmark      &        8         &     37          & ESO383-87            \\
        49923 &  14:01:21.6 & -33:03:49           & $29.09\pm0.20$    &                 &                 &        8         &     54          & N5398                \\
        50073 &  14:03:21.2 & -41:22:36           & $28.63\pm0.10$    &                 & \checkmark      &       10         &                 & N5408                \\
        51659 &  14:28:03.6 & -46:18:19           & $27.79\pm0.10$    &                 & \checkmark      &       10         &                 & PGC51659             \\
        62918 &  19:13:14.3 & -62:05:19           & $29.23\pm0.20$    &                 &                 &       10         &                 & I4824                \\
        63287 &  19:29:59.0 & -17:40:44           & $25.17\pm0.10$    &                 & \checkmark      &       10         &                 & Sag DIG              \\
        63616 &  19:44:57.0 & -14:48:01           & $23.41\pm0.06$    & \checkmark      & \checkmark      &       10         &                 & NGC6822              \\
        64054 &  20:03:57.3 & -31:40:54           & $29.18\pm0.10$    &                 & \checkmark      &       10         &                 & KK246                \\
        64181 &  20:09:31.7 & -61:51:02           & $29.56\pm0.20$    &                 &                 &        8         &     77          & I4951                \\
        65367 &  20:46:51.7 & -12:50:51           & $25.02\pm0.08$    &                 & \checkmark      &       10         &                 & Aquarius dIrr        \\
        67045 &  21:36:28.9 & -54:33:27           & $29.70\pm0.09$    &                 & \checkmark      &        5         &     78          & N7090                \\
        67908 &  22:02:41.4 & -51:17:48           & $26.46\pm0.10$    &                 & \checkmark      &       10         &                 & I5152                \\
        68672 &  22:22:30.5 & -48:24:14           & $29.52\pm0.10$    &                 & \checkmark      &       10         &                 & ESO238-005           \\
        70027 &  22:55:45.7 & -42:38:31           & $29.57\pm0.20$    &                 &                 &        3         &     43          & N7412                \\
        71431 &  23:26:27.9 & -32:23:19           & $26.72\pm0.08$    &                 & \checkmark      &       10         &                 & UGCA438              \\
        71866 &  23:36:15.0 & -37:56:19           & $29.86\pm0.20$    &                 &                 &        7         &     65          & N7713                \\
        72228 &  23:43:45.1 & -31:57:34           & $28.20\pm0.10$    &                 & \checkmark      &        9         &     78          & UGCA442              \\
        73049 &  23:57:49.8 & -32:35:28           & $27.77\pm0.07$    & \checkmark      & \checkmark      &        7         &     55          & N7793                
\enddata
\tablecomments{Distance modulus \& uncertainty and morphological type from \citet{tully13}. $\checkmark$ in columns labeled ``C?'' and ``T?'' denote existing Cepheid and TRGB distance determinations. Inclination values as reported by NED, based on the $B=26$~mag/\sq\arcsec isophote.}
\end{deluxetable}
\end{document}